\documentclass[aip,pof,preprint,amsmath,amssymb]{revtex4-2}

\usepackage{graphicx}
\usepackage[caption=false]{subfig}

\graphicspath{{./figures/}}

\begin{document}

\preprint{AIP/123-QED}

\title[HACS GSI model]{Rarefied Aerodynamics of Complex Satellites: A Unified Gas-Kinetic and Monte Carlo Framework}

\author{Vignesh Ram Petha Sethuraman}
\email{rampsv@sejong.ac.kr}
\affiliation{Department of Aerospace System Engineering, Sejong University, Seoul-05000, South Korea}

\author{Wu Kexin}
\affiliation{Zhejiang Key Laboratory of Multiflow and Fluid Machinery, Zhejiang Sci-Tech University, Hangzhou 310018, PR China}

\author{S Ajith Kumar}
\affiliation{Department of Mechanical Engineering, Amrita Vishwa Vidyapeetham, Amritapuri 690525, India}

\begin{abstract}
The aerodynamic loads and heating on a satellite in very low Earth orbit depend on which molecules reach its surfaces and how they scatter. We present HACS, a free-molecular solver with gas-kinetic and particle Monte Carlo formulations on a common triangulated geometry. The gas-kinetic formulation integrates molecular fluxes with thermally averaged molecular visibility (TAMV), while Monte Carlo traces individual first impacts. Both use Maxwell and Cercignani--Lampis--Lord (CLL) scattering and retain incident and reflected contributions to the local and integrated loads. Flat-plate tests check the incident fluxes and reflection models, and cube and open-box tests examine three-dimensional integration and shadowing. At the selected CLL flat-plate conditions, lift and drag from HACS agree with SPARTA DSMC within 1\% in relative terms, with the normal-incidence lift assessed in absolute terms because it vanishes by symmetry. On a JAXA SLATS model and a more complex satellite, gas-kinetic and Monte Carlo calculations give similar integrated trends over pitch and yaw sweeps at altitudes from 120 to 300~km. Compared with the bulk-direction visibility treatment, TAMV captures gradual pressure changes near shadow boundaries and grazing surfaces, where thermal motion allows molecules to reach partially obstructed regions.
\end{abstract}


\maketitle

\section{Introduction}\label{S:1}

Flying closer to Earth improves the spatial resolution attainable with a given optical aperture and reduces the range over which a communication link must operate. These advantages have renewed interest in very low Earth orbit (VLEO), often taken to extend up to about 400--450~km~\cite{crisp_2020,sabrina_2020}. JAXA's Super Low Altitude Test Satellite (SLATS, or Tsubame) demonstrated operation at progressively lower altitudes, reaching about 167~km. Its onboard atomic-oxygen monitor also measured the interaction of the upper atmosphere with exposed materials~\cite{kimoto_2022}. At orbital speeds of roughly $7$--$8~\mathrm{km\,s^{-1}}$, the residual atmosphere transfers appreciable momentum and energy to spacecraft surfaces. Drag affects orbital decay and the propellant required for orbit maintenance. The corresponding moments can disturb pointing or provide control authority when the spacecraft is designed to use aerodynamic actuation~\cite{geyer_2026}. Surface heating and the influence of geometry on loads also matter when choosing a configuration for sustained operation~\cite{hild_2022,walsh_2021}.

A constant drag coefficient such as $C_D=2.2$ can obscure changes due to spacecraft attitude, surface condition, atmospheric composition, and temperature~\cite{Moe_2005,mehta_2014comparing}. Thermospheric properties also vary with location, local solar time, and solar and geomagnetic activity. Empirical models such as NRLMSISE-00 provide species densities and temperature for a prescribed state~\cite{picone_2002nrlmsise}. Response surfaces offer a way to represent the resulting variation in satellite drag coefficients over a range of input conditions~\cite{mehta_2014}. The choice of flow model depends on the Knudsen number, $\mathrm{Kn}=\lambda/L$, where $\lambda$ is the mean free path and $L$ is a characteristic spacecraft dimension. In the free-molecular limit, loads follow from the momentum and energy exchanged at the surface~\cite{sentman_1961,Bird1994}. Panel methods apply this description to spacecraft geometries and account for geometric shading~\cite{fuller_2009,hart_2018}. When intermolecular collisions influence the flow around the spacecraft, direct simulation Monte Carlo (DSMC) can resolve their effect, as demonstrated in drag modeling for GRACE~\cite{mehta_2013}. The altitude range of this study, 120--300~km, can span different rarefaction levels depending on the atmospheric state and spacecraft size.

Gas--surface interaction (GSI) remains a source of uncertainty even in the free-molecular limit. Surface contamination and atomic-oxygen exposure can change the scattering behavior, and the magnitude of the effect depends on the surface and the incident species~\cite{Moe_2005,sabrina_2020}. Semiempirical and generalized energy-accommodation formulations describe aspects of this energy exchange~\cite{pilinski_2010,tuttas_2025}. The Satellite for Orbital Aerodynamics Research (SOAR) mission motivated an in-orbit method for characterizing drag-reducing materials~\cite{crisp_2022}. The Maxwell model combines specular and diffuse reflection. The specular branch reverses the normal component of velocity and preserves the tangential component. The diffuse branch emits molecules with a thermal flux distribution at the wall temperature~\cite{Maxwell1879,Bird1994}. The Cercignani--Lampis--Lord (CLL) model assigns independent parameters to normal energy and tangential momentum accommodation. The resulting scattering distribution can retain a preferred direction while spreading about it~\cite{CercignaniLampis1971,Lord1991}. Comparisons of GSI models show that the treatment of incomplete accommodation can affect predicted satellite drag coefficients~\cite{walker_2014,mehta_2014comparing,sabrina_2020}.

On a spacecraft with solar arrays, booms, or recesses, one component can shield another from the incoming gas. This changes where momentum and energy are transferred to the surface~\cite{fuller_2009,ramesh_2025}. Near a shadow boundary, a panel blocked along the mean freestream direction can still receive molecules whose thermal velocities carry them around the obstruction. The visibility of that panel therefore depends on the direction of the incident molecular velocity.

In the present work, the free-molecular gas-kinetic (GK) formulation of HACS is extended to calculate aerodynamic forces, moments, and surface heat flux for VLEO spacecraft at altitudes from 120 to 300~km. In the previous HACS formulation, molecules were assumed to undergo fully diffuse reflection with complete accommodation at the wall temperature~\cite{PethaSethuraman2024HACS}. Since incomplete accommodation can affect the momentum and energy exchanged at spacecraft surfaces~\cite{Moe_2005,walker_2014,sabrina_2020}, the Maxwell and CLL scattering models are incorporated into the present formulation. In the proposed analytical method, thermally averaged molecular visibility (TAMV) is used to account for geometric shadowing. The visibility of each surface element is evaluated for the molecular velocity directions used in the gas-kinetic integration. This allows molecules with transverse thermal velocities to contribute to the flux near shadow boundaries. A test-particle Monte Carlo (MC) method is also implemented using the same triangulated geometry, freestream conditions, and scattering models. In this method, molecular velocities are sampled and the trajectories are traced to their first surface impact~\cite{davis_1960,ramesh_2025}. The incident and reflected contributions are calculated separately in both methods to examine their effects on the local surface loads and the total forces, moments, and heating.

To assess the proposed method, calculations are first performed for flat plates, a cube, and an open box. The method is then applied to a model of the JAXA SLATS spacecraft and to a more complex satellite to examine the effects of altitude, attitude, and surface accommodation. The analytical results are compared with the MC results, and for selected conditions the HACS predictions are also compared with calculations from the SPARTA direct simulation Monte Carlo (DSMC) solver~\cite{Plimpton2019}.

\section{Rarefied Aerodynamic and Gas--Surface Interaction Formulation}
\label{sec:formulation}

In the body frame, HACS treats the triangulated spacecraft surface as stationary and evaluates the momentum and energy exchanged at its walls. The free-molecular model neglects intermolecular collisions near the body~\cite{Bird1994}. GK evaluates velocity moments, while MC estimates the same transfers from sampled first impacts.

\subsection{Freestream distribution and incident fluxes}
\label{sec:freestream_distribution}
\label{sec:incident_flux}

Let species $s$ have molecular mass $m_s$, number density $n_s$, and $\zeta_s$ internal quadratic degrees of freedom. Its velocity distribution is the drifting Maxwellian
\begin{equation}
 f_s(\mathbf v)=\frac{n_s\beta_s^3}{\pi^{3/2}}
 \exp[-\beta_s^2|\mathbf v-\mathbf U_\infty|^2],
 \qquad
 \beta_s=\sqrt{\frac{m_s}{2k_BT_\infty}} .
 \label{eq:drifting_maxwellian}
\end{equation}
On a panel with outward unit normal $\mathbf n$, an incoming molecule has inward speed $w=-\mathbf v\cdot\mathbf n>0$. Define $V_N=-\mathbf U_\infty\cdot\mathbf n$, $\mathbf V_T=\mathbf U_\infty+V_N\mathbf n$, $s_n=\beta_s V_N$, $s_t=\beta_s|\mathbf V_T|$, and $S_s^2=s_n^2+s_t^2$. The signed $s_n$ describes bulk incidence, while the individual condition $w>0$ selects the incident half range.

On an unobstructed panel, let $M_j$ be the dimensional half-range moments of the normal speed from Appendix~\ref{app:incident_flux}. With $\rho_s=m_sn_s$, the incident number flux, pressure, shear magnitude, and energy flux are
\begin{flalign}
 &\Gamma_s = n_s M_1, && \label{eq:incident_number_flux}\\
 &p_{i,s} = \rho_s M_2, && \label{eq:incident_pressure}\\
 &\tau_{i,s} = \rho_s M_1|\mathbf V_T|, && \label{eq:incident_shear}\\
 &q_{i,s} = \frac{\rho_s}{2}
 \left[M_3+\frac{M_1}{\beta_s^2}
 \left(s_t^2+1+\frac{\zeta_s}{2}\right)\right]. &&
 \label{eq:incident_energy}
\end{flalign}
The shear vector points along $\mathbf V_T$ and is zero when $\mathbf V_T=\mathbf 0$. At a given velocity, the incident energy per molecule is $e_i=m_s|\mathbf v|^2/2+\zeta_s k_B T_\infty/2$. Section~\ref{sec:mc_exchange} uses the same energy in the impact-rate estimator.

\subsection{Reflected momentum and energy}
\label{sec:reflected_flux}

\begin{figure}[htbp]
 \centering
 \includegraphics[width=\linewidth]{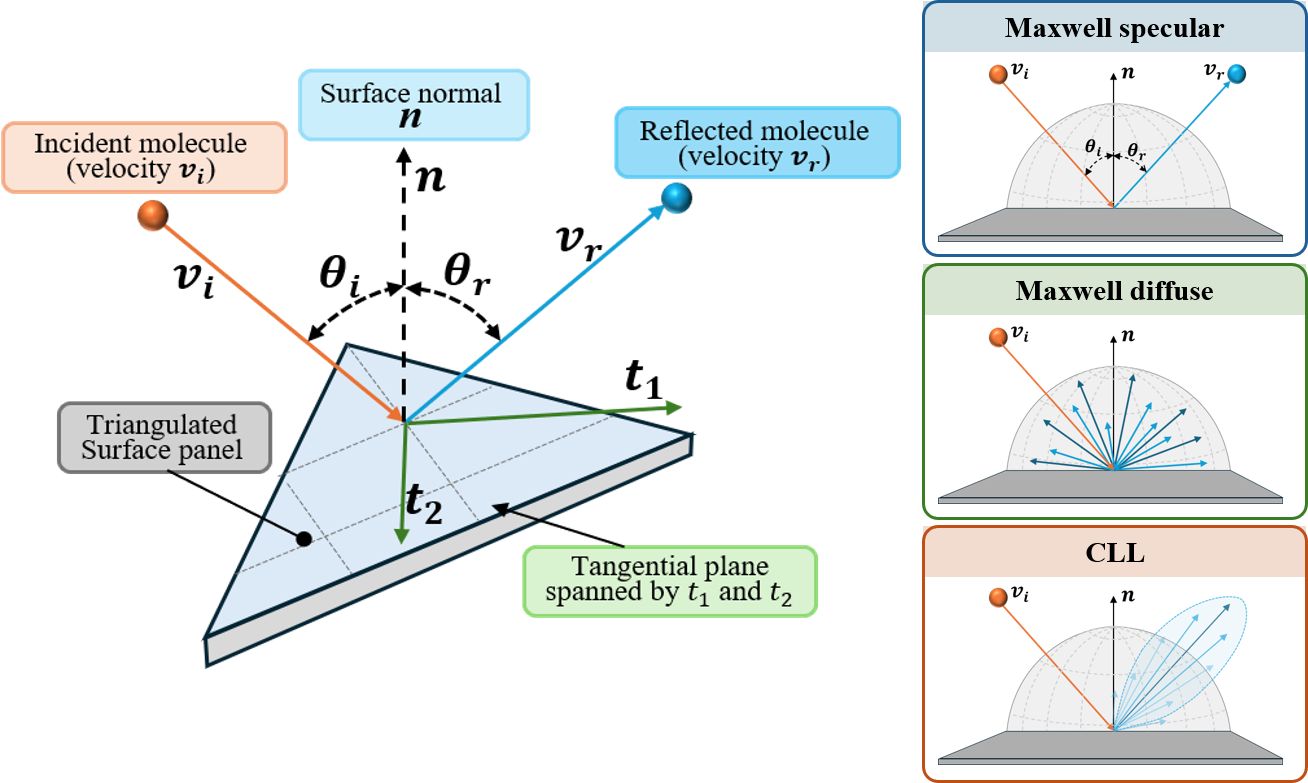}
 \caption{Incident and reflected molecular velocities on a triangular panel. The local frame consists of the outward normal and two tangential directions. The sketches show specular Maxwell, diffuse Maxwell, and directional CLL scattering.}
 \label{fig:gsi_schematic}
\end{figure}

The GK scales for pressure and shear are $n_s k_BT_\infty$, and the scale for heat flux is $n_s k_BT_\infty/\beta_s$. Positive pressure denotes compression. The reflected shear coefficient carries the sign of its contribution to wall traction. The reflected energy coefficient $C_{h,r}$ measures outgoing energy, giving net heating $C_{h,i}-C_{h,r}$. Figure~\ref{fig:gsi_schematic} shows the surface frame and the scattering models.

\subsubsection{Maxwell reflection}
\label{sec:maxwell_gsi}

Specular reflection reverses the molecular normal velocity and preserves tangential velocity and internal energy. Diffuse reflection emits molecules according to the outgoing thermal flux distribution at wall temperature $T_w$ \cite{Bird1994}. Define
\[
 E_n=e^{-s_n^2},\quad R_n=1+\operatorname{erf}(s_n),\quad
 A_n=E_n+\sqrt{\pi}s_nR_n,\quad t_r=T_w/T_\infty .
\]
In GK, $\sigma_N$, $\sigma_T$, and $\sigma_E$ are in $[0,1]$ and separately interpolate the specular and diffuse limits for normal momentum, tangential momentum, and total energy. The resulting coefficients are
\begin{align}
 C_{p,r}^{M} &=(1-\sigma_N)C_{p,i}+\tfrac12\sigma_N\sqrt{t_r}A_n,
 \label{eq:maxwell_pressure_ref}\\
 C_{t,r}^{M} &=-(1-\sigma_T)C_{t,i},\qquad
 C_t^{M}=\sigma_T C_{t,i}, \label{eq:maxwell_shear}\\
 C_{h,r}^{M} &=(1-\sigma_E)C_{h,i}
      +\sigma_E\frac{(\zeta_s+4)t_r A_n}{4\sqrt{\pi}} .
 \label{eq:maxwell_heat_ref}
\end{align}
Appendix~\ref{app:cll_derivation} derives these coefficients from the conditional reflected moments. MC draws a single diffuse or specular event with diffuse probability $\sigma_N$. Its mean transfers therefore coincide with the GK rules when $\sigma_N=\sigma_T=\sigma_E$.

\subsubsection{Cercignani--Lampis--Lord reflection}
\label{sec:cll_gsi}

HACS parameterizes the CLL kernel by the normal energy accommodation $\alpha_N\in[0,1]$ and the tangential momentum accommodation $\alpha_T\in[0,2]$~\cite{CercignaniLampis1971,Lord1991}. Here $\alpha_E\in[0,1]$ controls the mean internal energy, and $a_T=\alpha_T(2-\alpha_T)$ is the tangential translational energy accommodation. Define $\beta_w=\sqrt{m_s/(2k_BT_w)}$ and the wall-scaled normal speeds $c_i=\beta_w w$ and $c_r=\beta_w v_{r,n}$. For $\alpha_N>0$ the conditional normal kernel is
\begin{equation}
 K_n(c_r\mid c_i)=\frac{2c_r}{\alpha_N}
 \exp\!\left[-\frac{c_r^2+(1-\alpha_N)c_i^2}{\alpha_N}\right]
 I_0\!\left(\frac{2\sqrt{1-\alpha_N}\,c_i c_r}{\alpha_N}\right),
 \quad (c_r>0).
 \label{eq:cll_kernel}
\end{equation}

Write $\mathcal B(z)=(1+2z)I_{0e}(z)+2zI_{1e}(z)$, where $I_{\nu e}(z)=e^{-z}I_\nu(z)$ and $I_\nu$ is the modified Bessel function of the first kind. For conventional GK the exact normal-moment integral and the fast approximation are
\begin{align}
 C_{p,r}^{\mathrm{exact}}
 &=\sqrt{\alpha_N t_r}\int_0^\infty x e^{-(x-s_n)^2}
             \mathcal B\bigl(z(x)\bigr)\,dx,
 &z(x)&=\frac{(1-\alpha_N)x^2}{2\alpha_N t_r},
 \label{eq:cll_exact_pressure}\\
 C_{p,r}^{\mathrm{fast}}
 &=\frac{A_n}{2}\sqrt{\alpha_N t_r}\,
        \mathcal B\bigl(z(\beta_s M_2/M_1)\bigr).
 \label{eq:cll_fast_pressure}
\end{align}
CLL-exact integrates the normal moment using composite Simpson quadrature with 1000 subintervals over $[0,\max(8,s_n+8)]$ and piecewise approximations to the scaled Bessel functions. CLL-fast evaluates the same conditional mean at the impact-weighted speed $M_2/M_1$. Both branches set $C_{p,r}=C_{p,i}$ for $\alpha_N\le10^{-10}$. Above this cutoff, CLL-fast sets $C_{p,r}=0$ when $A_n/\sqrt{\pi}$ is numerically zero.

Both conventional pressure branches use the same shear and energy expressions. With $B_n=\beta_s^3M_3$,
\begin{align}
 C_{t,r}^{\mathrm{CLL}} &=-(1-\alpha_T)C_{t,i},\qquad
 C_t^{\mathrm{CLL}}=\alpha_T C_{t,i}, \label{eq:cll_shear}\\
 C_{h,r,\mathrm{tr}}^{\mathrm{CLL}}
 &=(1-\alpha_N)B_n+\frac{A_n}{2\sqrt{\pi}}
 \left[(1-\alpha_T)^2(s_t^2+1)+(\alpha_N+a_T)t_r\right],
 \label{eq:cll_heat_trans}\\
 C_{h,r,\mathrm{int}}^{\mathrm{CLL}}
 &=(1-\alpha_E)C_{h,i,\mathrm{int}}
   +\alpha_E\frac{\zeta_s t_r A_n}{4\sqrt{\pi}} .
 \label{eq:cll_heat_internal}
\end{align}
The translational and internal terms add to give the outgoing energy flux. GK--TAMV uses the conditional mean velocity and energy at each quadrature node for both CLL selectors. MC samples the CLL translational kernel and assigns the same mean accommodated internal energy.

\section{HACS Computational Framework}
\label{sec:hacs_framework}

\subsection{Inputs and conventional gas-kinetic calculations}
\label{sec:hacs_inputs}
\label{sec:multispecies}
\label{sec:hacs_workflow}
\label{sec:gk_solver}

\begin{figure}[htbp]
 \centering
 \includegraphics[width=\linewidth]{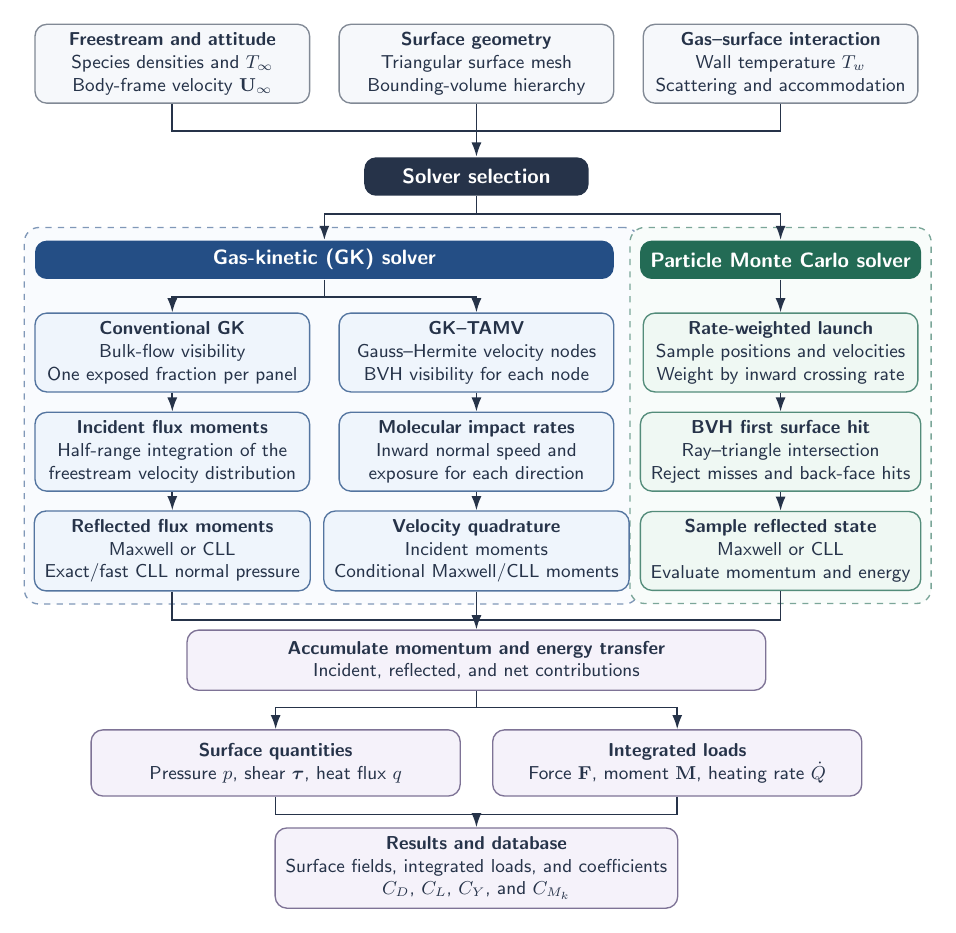}
 \caption{HACS workflow from species-resolved atmospheric and surface inputs to incident, reflected, and net loads and heating.}
 \label{fig:hacs_architecture}
\end{figure}

The inputs to HACS are the freestream state, spacecraft attitude, triangulated panel geometry, wall temperatures, and accommodation coefficients. Each species is evaluated using its own mass, internal degrees of freedom, and number density. If composition is supplied as fractions, the solver obtains the species number density by multiplying the total number density by the corresponding number fraction. Figure~\ref{fig:hacs_architecture} shows the workflow.

Conventional GK assigns a single exposed-area fraction $\chi_a^{\mathrm{conv}}=1-\mathrm{shade\_area}_a$ to each panel using the bulk-flow direction. The solver skips fully shadowed panels. For the remaining panels, the same fraction multiplies the sum of species fluxes.
\begin{align}
 \mathbf F_{j,a}^{\mathrm{conv}}
 &= A_a\chi_a^{\mathrm{conv}}\sum_s
     (-p_{j,s}\mathbf n_a+\boldsymbol{\tau}_{j,s}),
 \qquad j\in\{i,r\}, \label{eq:element_force}\\
 \dot Q_{\mathrm{net},a}^{\mathrm{conv}}
 &= A_a\chi_a^{\mathrm{conv}}\sum_s(q_{i,s}-q_{r,s}).
 \label{eq:species_sum}
\end{align}
The half-range moments retain the signed incidence parameter $s_n$. The reflected shear is a signed wall contribution, and $q_{r,s}$ is the outgoing energy flux.

\subsection{Velocity-dependent exposure in GK--TAMV}
\label{sec:surface_shadowing}

TAMV integrates over molecular directions using a tensor-product physicists' Gauss--Hermite (GH) rule~\cite{DLMFQuadrature}, with order $N_{\mathrm{GH}}\in\{5,7,9,11\}$ in each Cartesian component. The velocity and normalized weight of node $q=(i,j,k)$ are
\[
 \mathbf v_{s,q}=\mathbf U_\infty+\beta_s^{-1}(x_i,x_j,x_k),
 \qquad \omega_q=\pi^{-3/2}w_iw_jw_k .
\]
A node contributes when $v_{n,a,s,q}=-\mathbf v_{s,q}\cdot\mathbf n_a>0$. The bounding-volume hierarchy (BVH) checks for obstructions along the upstream direction $-\mathbf v_{s,q}/|\mathbf v_{s,q}|$. Depending on the panel flag, the solver assigns full exposure or full shadow, tests a centroid ray, or checks three rays and subtracts the blocked fraction. With $\chi_{a,s,q}$ denoting the resulting exposed fraction, the impact rate at the node is
\begin{equation}
 \dot N_{a,s,q}=n_s A_a\,\omega_q v_{n,a,s,q}\chi_{a,s,q}.
 \label{eq:tamv_rate}
\end{equation}
That rate weights both incident and reflected molecular properties.
\begin{align}
 \mathbf F_{i,a}&=\sum_{s,q}\dot N_{a,s,q}m_s\mathbf v_{s,q},&
 \mathbf F_{r,a}&=-\sum_{s,q}\dot N_{a,s,q}m_s\overline{\mathbf v}_{r,a,s,q},\\
 \dot Q_{i,a}&=\sum_{s,q}\dot N_{a,s,q}e_{i,s,q},&
 \dot Q_{r,a}^{\mathrm{out}}&=\sum_{s,q}\dot N_{a,s,q}\overline e_{r,a,s,q}.
 \label{eq:tamv_exchange}
\end{align}
The sums include only incoming nodes. Appendix~\ref{app:cll_derivation} gives the conditional reflected moments used at those nodes. For CLL, the TAMV normal mean takes its specular value when $\alpha_N\le10^{-12}$. The reflected energy uses the prescribed accommodation coefficients throughout.

The reported TAMV visibility is a ratio of incident number fluxes:
\begin{equation}
 \chi_a^*=\frac{\sum_s n_s\sum_q\omega_q v_{n,a,s,q}\chi_{a,s,q}}
                         {\sum_s n_s\sum_q\omega_q v_{n,a,s,q}}.
 \label{eq:tamv_visibility}
\end{equation}
The solver returns zero when its denominator is numerically zero. Forces and heat rates incorporate directional visibility through Eq.~\eqref{eq:tamv_rate}. Appendix~\ref{app:tamv_derivation} derives both the rate and the diagnostic from the velocity integral.

\subsection{Particle Monte Carlo}
\label{sec:mc_solver}

For each active species, HACS samples velocities from $f_s/n_s$ and launch positions uniformly by area on an upstream rectangle or an enclosing sphere. If $\mathbf n_{\mathrm{in}}$ is the inward launch normal, the sample is retained for tracing when $v_{\mathrm{cross}}=\mathbf v_i\cdot\mathbf n_{\mathrm{in}}>0$. Its physical molecular rate is
\begin{equation}
 W_p=\frac{n_s A_{\mathrm{launch}}}{N_{\mathrm{MC}}}\,
                    v_{\mathrm{cross}}.
 \label{eq:mc_launch_rate}
\end{equation}
Here $N_{\mathrm{MC}}$ counts launch attempts for each species. The BVH finds the nearest geometric intersection using a ray--triangle test~\cite{MollerTrumbore1997}. Misses and back-facing first intersections contribute zero. Each accepted trajectory supplies one pair of incident and reflected transfers at the first hit.

\paragraph{Maxwell sampling.}
The specular branch occurs with probability $1-\sigma_N$ and uses $\mathbf v_r=\mathbf v_i+2w\mathbf n$ and $e_{r,\mathrm{int}}=\zeta_s k_BT_\infty/2$. The diffuse branch occurs with probability $\sigma_N$. In an orthonormal surface basis it draws
\[
 v_{r,n}=\beta_w^{-1}\sqrt{-\ln U},\qquad
 v_{r,t_j}=\frac{Z_j}{\sqrt{2}\beta_w},\quad j=1,2,
\]
where $U$ is uniform on $(0,1)$ and $Z_1,Z_2$ are independent standard normal variables. The diffuse internal energy is set to its wall-temperature mean $\zeta_s k_BT_w/2$.

\paragraph{CLL sampling.}
Let $c_{i,t_j}=\beta_w v_{i,t_j}$. For four independent standard normal variables $Z_1,\ldots,Z_4$, the reflected wall-scaled speeds are
\begin{align}
 c_r &= \left[
 \left(\sqrt{1-\alpha_N}c_i+\sqrt{\alpha_N/2}Z_1\right)^2
 +\frac{\alpha_N}{2}Z_2^2\right]^{1/2}, \label{eq:cll_normal_sampling}\\
 c_{r,t_j} &=(1-\alpha_T)c_{i,t_j}+\sqrt{a_T/2}Z_{j+2},
 \qquad j=1,2 . \label{eq:cll_tangential_sampling}
\end{align}
Scaling by $1/\beta_w$ recovers the dimensional velocity. The driver uses $e_{r,\mathrm{int}}=(\zeta_s k_B/2)[(1-\alpha_E)T_\infty+\alpha_E T_w]$.

\paragraph{Momentum and energy exchange.}
\label{sec:mc_exchange}
For each accepted hit, the signed transfers to the wall are
\begin{align}
 \Delta\mathbf p_i&=m_s\mathbf v_i,&
 \Delta\mathbf p_r&=-m_s\mathbf v_r,\label{eq:mc_momentum}\\
 \Delta E_i&=\tfrac12m_s|\mathbf v_i|^2+\tfrac{\zeta_s}{2}k_BT_\infty,&
 \Delta E_r&=-\bigl(\tfrac12m_s|\mathbf v_r|^2+e_{r,\mathrm{int}}\bigr).
 \label{eq:mc_energy}
\end{align}
Multiplying each transfer by $W_p$ gives its contribution to the panel force or heating rate. The moment arm runs from the reference origin to the actual hit point. A separate visibility calculation estimates Eq.~\eqref{eq:tamv_visibility} from sampled velocities and directional BVH exposure, with equal sampling weights within each species. Appendix~\ref{app:mc_estimators} derives the launch weights and both estimators.

\subsection{Surface and integrated quantities}
\label{sec:hacs_outputs}
\label{sec:force_moment}

Dividing panel force by area gives the traction $\mathbf f_{j,a}=\mathbf F_{j,a}/A_a$. Its pressure and shear components are $p_{j,a}=-\mathbf f_{j,a}\cdot\mathbf n_a$ and $\boldsymbol{\tau}_{j,a}=\mathbf f_{j,a}+p_{j,a}\mathbf n_a$. Heat flux is the signed heating rate divided by panel area. HACS records the reflected energy leaving the surface as a negative contribution, so summing the signed incident and reflected terms gives the net heat entering the wall.

The spacecraft force and heating rate are the sums over panels. GK evaluates each moment about $\mathbf x_0$ with the panel centroid $\mathbf x_a$:
\begin{equation}
 \mathbf F_j=\sum_a\mathbf F_{j,a},\qquad
 \mathbf M_j^{\mathrm{GK}}=\sum_a(\mathbf x_a-\mathbf x_0)\times\mathbf F_{j,a},
 \qquad \dot Q_j=\sum_a \dot Q_{j,a}.
 \label{eq:force_moment_total}
\end{equation}
MC sums $\bigl(\mathbf x_{\mathrm{hit},p}-\mathbf x_0\bigr)\times W_p\Delta\mathbf p_{j,p}$ over accepted hits. The drag basis vector follows the freestream direction, lift uses the projected reference-up direction with a fallback near parallel alignment, and the side direction is $\mathbf e_L\times\mathbf e_D$. Thus $(D,L,Y)=(\mathbf F\cdot\mathbf e_D,\mathbf F\cdot\mathbf e_L,\mathbf F\cdot\mathbf e_S)$. With $\rho_\infty=\sum_s\rho_s$ and $q_\infty=\rho_\infty|\mathbf U_\infty|^2/2$, the coefficient definitions are
\begin{equation}
 (C_D,C_L,C_Y)=\frac{(D,L,Y)}{q_\infty A_{\mathrm{ref}}},
 \qquad
 C_{M_k}=\frac{M_k}{q_\infty A_{\mathrm{ref}}L_{\mathrm{ref}}}.
 \label{eq:force_coefficients}
\end{equation}

\subsection{Solver verification and DSMC benchmarking}
\label{sec:analytical_mc_consistency}

The incident terms isolate incidence, visibility, and rate weighting. The reflected terms also depend on accommodation. For an unobstructed geometry and matched scattering moments, the conventional GK half-range result provides a reference for MC. On a complex body, conventional GK uses a bulk-direction exposed fraction, GK--TAMV weights the directions represented in its quadrature, and MC traces individual impacts. Comparing conventional CLL-fast with CLL-exact isolates the representative-speed approximation in reflected pressure.

SPARTA provides the independent DSMC results used in the flat-plate benchmark and selected spacecraft comparisons~\cite{Bird1994,Plimpton2019}. The comparisons use the corresponding freestream, attitude, wall temperature, scattering settings, and load definitions. Differences can arise from spatial and statistical resolution and, when intermolecular collisions are enabled, from the flow physics represented by DSMC.

\section{Verification and DSMC Benchmarking}
\label{sec:validation}

\subsection{Flat-Plate Momentum-Transfer Verification}
\label{sec:flat_plate_verification}

The verification starts with a flat plate to examine molecular fluxes and surface scattering without geometric obstruction. A cube then tests integration across panels with different orientations, and an open box tests self-shadowing. A published CLL flat-plate benchmark provides the conditions for an independent SPARTA comparison.

\begin{figure*}[htbp]
\centering

\begin{minipage}[t]{0.32\textwidth}
\centering
\includegraphics[width=\linewidth]{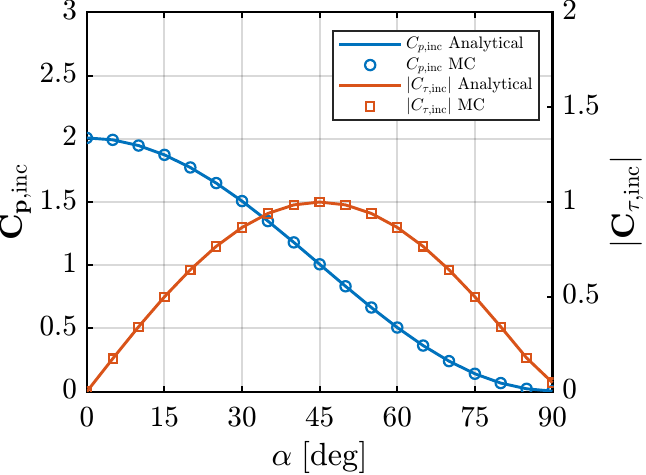}

{\small (a)}
\end{minipage}
\hfill
\begin{minipage}[t]{0.32\textwidth}
\centering
\includegraphics[width=\linewidth]{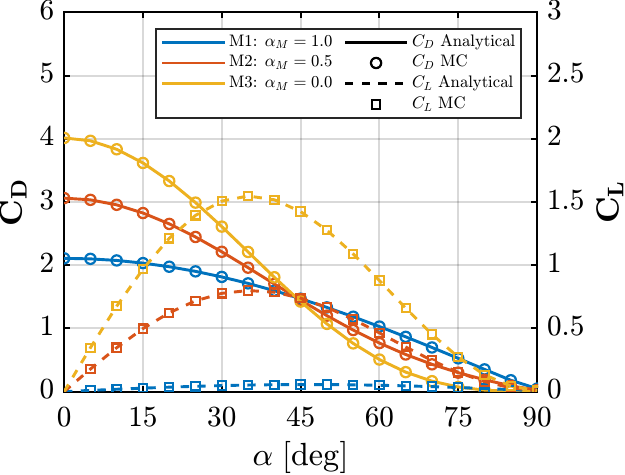}

{\small (b)}
\end{minipage}
\hfill
\begin{minipage}[t]{0.32\textwidth}
\centering
\includegraphics[width=\linewidth]{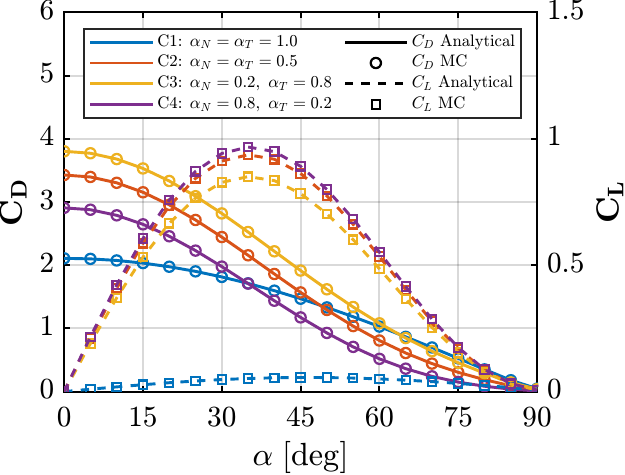}

{\small (c)}
\end{minipage}

\vspace{0.3cm}

\begin{minipage}[t]{0.32\textwidth}
\centering
\includegraphics[width=\linewidth]{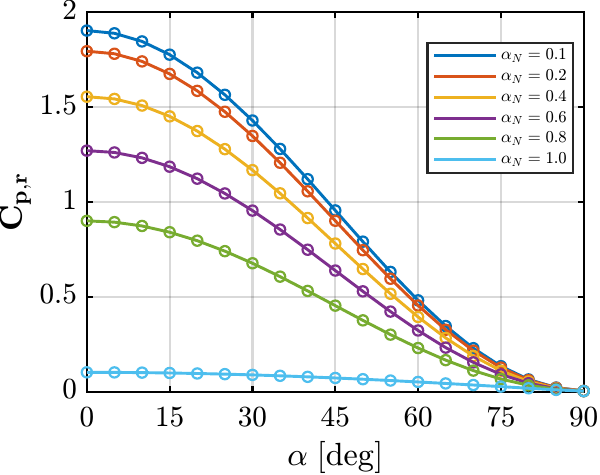}

{\small (d)}
\end{minipage}
\hspace{0.05\textwidth}
\begin{minipage}[t]{0.32\textwidth}
\centering
\includegraphics[width=\linewidth]{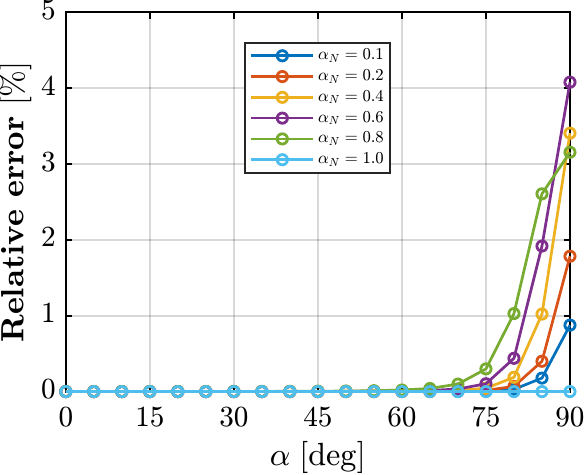}

{\small (e)}
\end{minipage}

\caption{Flat-plate comparisons at 150~km. Panel (a) shows the incident pressure and tangential momentum coefficients. Panels (b) and (c) show drag and lift for Maxwell and CLL reflection, respectively. Lines in (a)--(c) give the GK results and symbols give the MC results. Panel (d) compares CLL-exact (lines) and CLL-fast (symbols) for reflected pressure at $\alpha_N=0.1$--$1.0$ and $\alpha_T=1.0$. Panel (e) shows their relative difference.}
\label{fig:flat_plate_verification}
\end{figure*}

For the in-house flat-plate tests, the altitude is 150~km, the freestream speed is $7814~\mathrm{m\,s^{-1}}$, and $T_w=300$~K. Both solvers use the same geometry, reference area, and species-resolved state for O, N$_2$, O$_2$, He, Ar, H, N, and O$^*$. The angle $\alpha$ is measured from the normal. The sweep spans $0^\circ$ (normal incidence) to $90^\circ$ (grazing incidence). Figure~\ref{fig:flat_plate_verification}(a) isolates the incident coefficients. The mean relative difference in $C_{p,\mathrm{inc}}$ is $0.0323\%$, with a maximum of $0.2355\%$ at $85^\circ$. Omitting normal incidence, where the analytical tangential momentum coefficient vanishes, the mean difference in $C_{\tau,\mathrm{inc}}$ is $0.0240\%$ and the maximum is $0.1198\%$ at $85^\circ$. These comparisons check the species-resolved velocity sampling and incident flux weighting.

Figure~\ref{fig:flat_plate_verification}(b) compares the Maxwell models at $\alpha_M=1.0$, $0.5$, and $0.0$ (M1--M3), where $\alpha_M=\sigma_N=\sigma_T=\sigma_E$. From fully diffuse to specular reflection, the mean relative differences in $C_D$ are $0.0212\%$, $0.0253\%$, and $0.0144\%$, and the maxima are $0.1213\%$, $0.1402\%$, and $0.0335\%$. Mean lift differences are $0.1788\%$, $0.0560\%$, and $0.0145\%$. The largest lift discrepancy is $1.4828\%$ at $5^\circ$ in the diffuse case, where lift is small. The maxima for M2 and M3 are $0.1452\%$ and $0.0335\%$.

The CLL comparison in Fig.~\ref{fig:flat_plate_verification}(c) covers four $(\alpha_N,\alpha_T)$ pairs. These are $(1.0,1.0)$, $(0.5,0.5)$, $(0.2,0.8)$, and $(0.8,0.2)$. The last two pairs test unequal normal and tangential accommodation. In the order listed, the mean relative differences in $C_D$ are $0.0085\%$, $0.0111\%$, $0.0101\%$, and $0.0131\%$, with all maxima below $0.046\%$. Excluding angles at which the reference lift is close to zero, the mean relative differences in $C_L$ are $0.1335\%$, $0.0243\%$, $0.0274\%$, and $0.0204\%$. In the first case, the largest relative lift difference is $0.344\%$. The agreement checks the normal pressure integral against independently sampled CLL velocities.

Figures~\ref{fig:flat_plate_verification}(d) and (e) isolate the change in reflected normal pressure introduced by CLL-fast. For $\alpha_T=1.0$, the mean relative differences in $C_{p,r}$ are $0.058\%$, $0.120\%$, $0.247\%$, $0.348\%$, and $0.384\%$ for $\alpha_N=0.1$, $0.2$, $0.4$, $0.6$, and $0.8$, respectively. For $\alpha_N=1.0$, both branches agree to numerical precision because the conditional normal mean is independent of the incident speed.

The largest relative pressure difference, about $4.08\%$, occurs at grazing incidence for $\alpha_N=0.6$. The reference pressure coefficient there is about $2.94\times10^{-3}$, and the absolute difference is $1.20\times10^{-4}$. The largest percentage difference therefore occurs where the reflected pressure is small.

\subsection{Reflected Molecular Angular Distribution}
\label{sec:reflection_lobe}
Integrated coefficients constrain moments of the velocity distribution. To examine the distribution itself, $10^6$ reflected molecular velocities were generated per incidence condition for atomic oxygen at $7814~\mathrm{m\,s^{-1}}$ and $T_w=300$~K. The incident angles are $\theta_i=30^\circ$, $45^\circ$, and $60^\circ$, measured between the incoming direction and the inward normal, equivalently between the reversed incoming direction and the outward normal.

The reflected molecular direction is described using the signed scattering-plane angle
\begin{equation}
\theta_r^s =
\operatorname{atan2}
\left(
\mathbf{v}_r\cdot\mathbf{t},
\mathbf{v}_r\cdot\mathbf{n}
\right),
\label{eq:signed_reflection_angle}
\end{equation}
where $\mathbf{n}$ is the outward surface normal and $\mathbf{t}$ denotes the tangential direction contained within the incident plane. The signed angle spans $-90^\circ\leq\theta_r^s\leq90^\circ$ and therefore distinguishes reflected molecules traveling in opposite tangential directions. The distribution is obtained after integrating over the remaining out-of-plane velocity component.

For fully diffuse Maxwell reflection, the signed projected-angle probability density is
\begin{equation}
P(\theta_r^s)
=
\frac{1}{2}\cos\theta_r^s,
\qquad
-\frac{\pi}{2}
\leq
\theta_r^s
\leq
\frac{\pi}{2},
\label{eq:diffuse_signed_angle_pdf}
\end{equation}

\begin{figure*}[htbp]
\centering
\begin{minipage}[t]{0.32\textwidth}
\centering
\includegraphics[width=\linewidth]{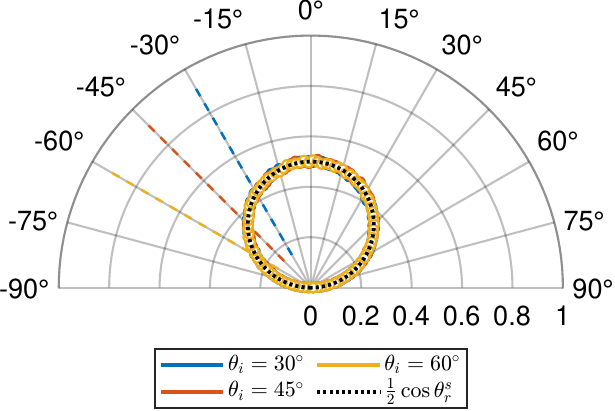}

{\small (a)}
\end{minipage}
\hfill
\begin{minipage}[t]{0.32\textwidth}
\centering
\includegraphics[width=\linewidth]{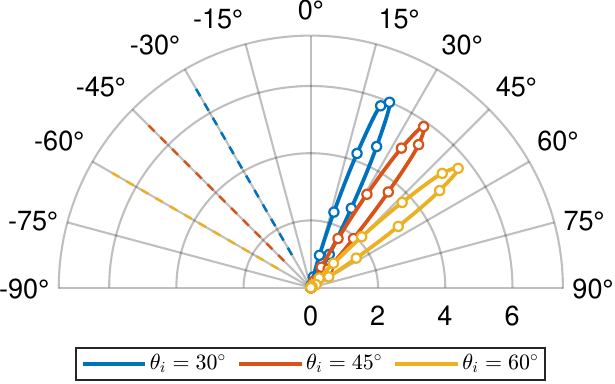}

{\small (b)}
\end{minipage}
\hfill
\begin{minipage}[t]{0.32\textwidth}
\centering
\includegraphics[width=\linewidth]{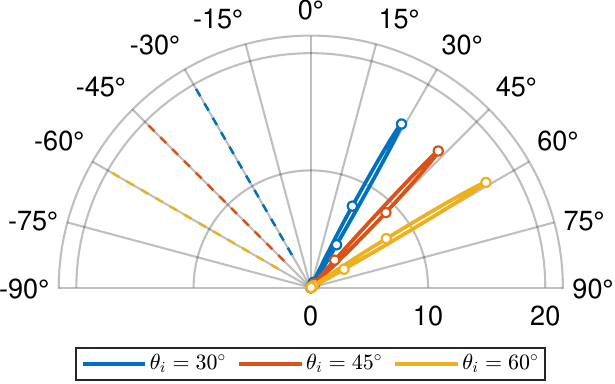}

{\small (c)}
\end{minipage}

\caption{Signed reflected-angle distributions for atomic oxygen at $\theta_i=30^\circ$, $45^\circ$, and $60^\circ$. Panel (a) uses diffuse Maxwell reflection ($\alpha_M=1$), and panels (b) and (c) use CLL with $\alpha_N=\alpha_T=0.5$ and $0.1$, respectively. Each distribution uses $10^6$ samples. Dashed radial lines mark the incident angles. Panel (a) includes the analytical projected density from Eq.~\eqref{eq:diffuse_signed_angle_pdf}.}
\label{fig:reflection_lobe}
\end{figure*}

The angle is measured in radians. Appendix~\ref{app:cll_derivation} derives the projected law from the diffuse velocity distribution. The three sampled distributions in Fig.~\ref{fig:reflection_lobe}(a) overlap, are symmetric about zero, and follow the cosine curve.

At $\alpha_N=\alpha_T=0.5$, the CLL peak shifts to larger positive angles as the incident direction becomes more oblique [Fig.~\ref{fig:reflection_lobe}(b)]. The model retains part of the incident tangential momentum, while the finite accommodation spreads the outgoing velocities~\cite{CercignaniLampis1971,Lord1991}. With $\alpha_N=\alpha_T=0.1$, the lobe narrows and its peak lies closer to the specular direction [Fig.~\ref{fig:reflection_lobe}(c)].

\subsection{Three-Dimensional Geometry and Shadowing Verification}
\label{sec:geometry_shadowing}

The convex cube tests surface orientation and force integration. The open box adds a recessed cavity, where parts of the surface are shielded by the rim or walls. Both tests use diffuse Maxwell reflection.

\begin{figure*}[htbp]
\centering

\begin{minipage}[t]{0.35\textwidth}
\centering
\includegraphics[width=\linewidth]{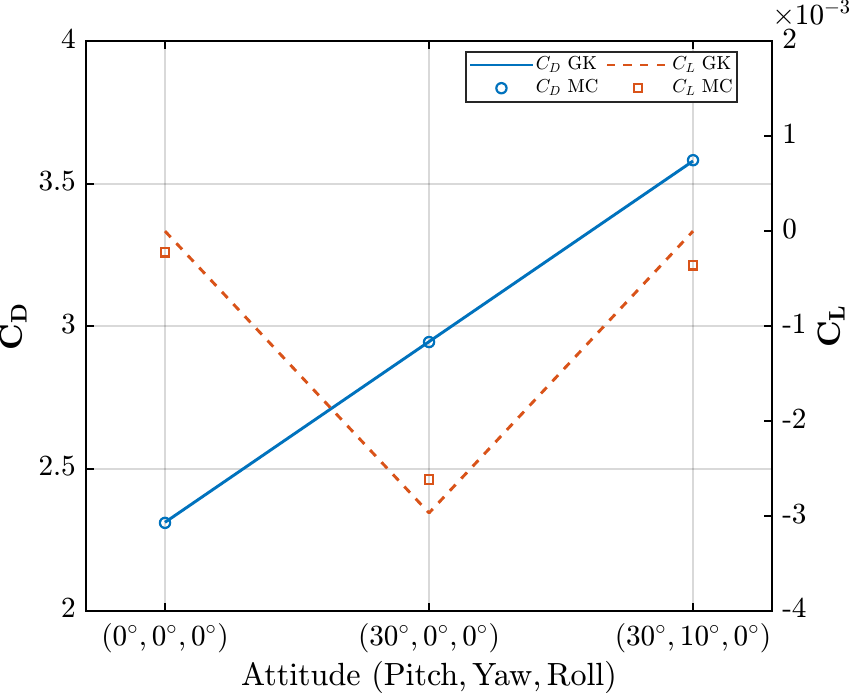}

{\small (a)}
\end{minipage}
\hfill
\begin{minipage}[t]{0.48\textwidth}
\centering
\includegraphics[width=\linewidth]{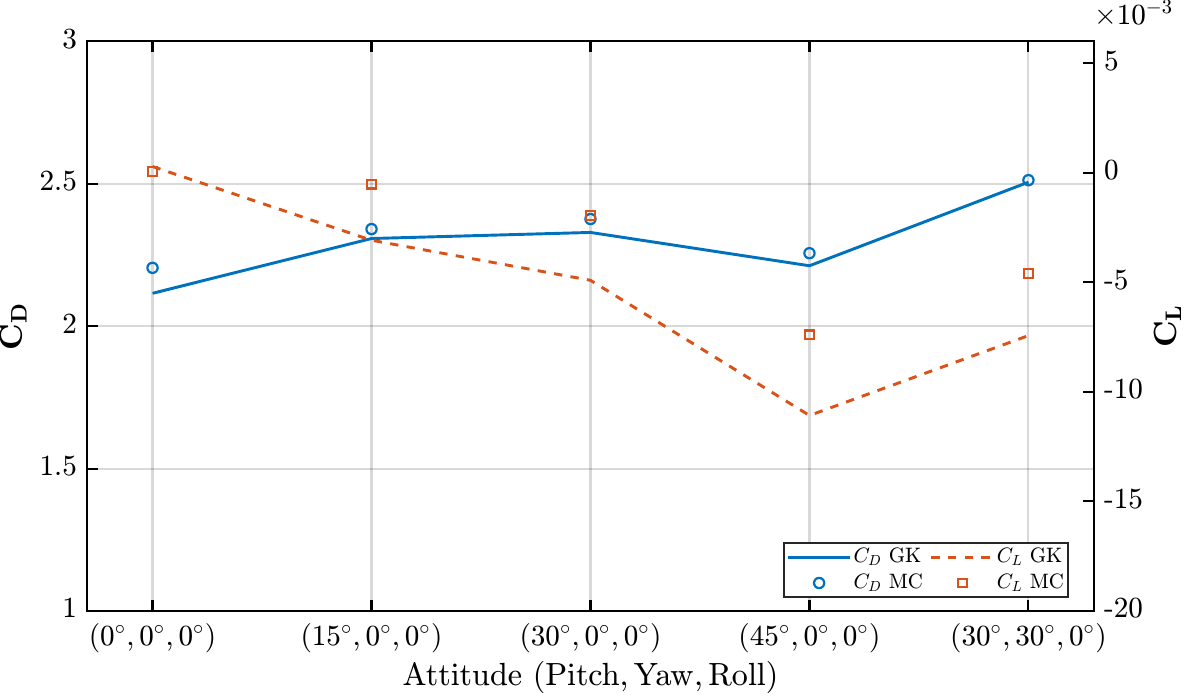}

{\small (b)}
\end{minipage}

\caption{Drag and lift coefficients under diffuse Maxwell reflection for (a) the cube and (b) the open box. Lines denote GK results and symbols denote MC results.}
\label{fig:geometry_verification}
\end{figure*}

The cube has a side length of 1~m and 12 triangular panels. The tested attitudes in pitch, yaw, and roll order are $(0^\circ,0^\circ,0^\circ)$, $(30^\circ,0^\circ,0^\circ)$, and $(30^\circ,10^\circ,0^\circ)$. At all three attitudes, GK and MC give nearly coincident $C_D$ and $C_L$ [Fig.~\ref{fig:geometry_verification}(a)]. Drag increases as more faces contribute to the projected area, and the lift remains small. 
The open box measures $1.2\times1.2\times0.6~\mathrm{m}$, with a $1.0\times1.0~\mathrm{m}$ opening, a cavity depth of 0.5~m, and a wall thickness of 0.1~m. The reference area is $1.44~\mathrm{m^2}$ and the reference length is 1.2~m. The pitch-only cases use $0^\circ$, $15^\circ$, $30^\circ$, and $45^\circ$. The oblique case, O3D, uses $(30^\circ,30^\circ,0^\circ)$.

The solvers give similar drag and lift at all five open-box orientations [Fig.~\ref{fig:geometry_verification}(b)]. The drag changes moderately as the projected box and opening change. The strongest negative lift occurs at $45^\circ$ pitch, where the cavity walls are exposed asymmetrically.
\begin{figure*}[htbp]
\centering

\begin{minipage}[t]{0.48\textwidth}
\centering
\includegraphics[width=\linewidth]{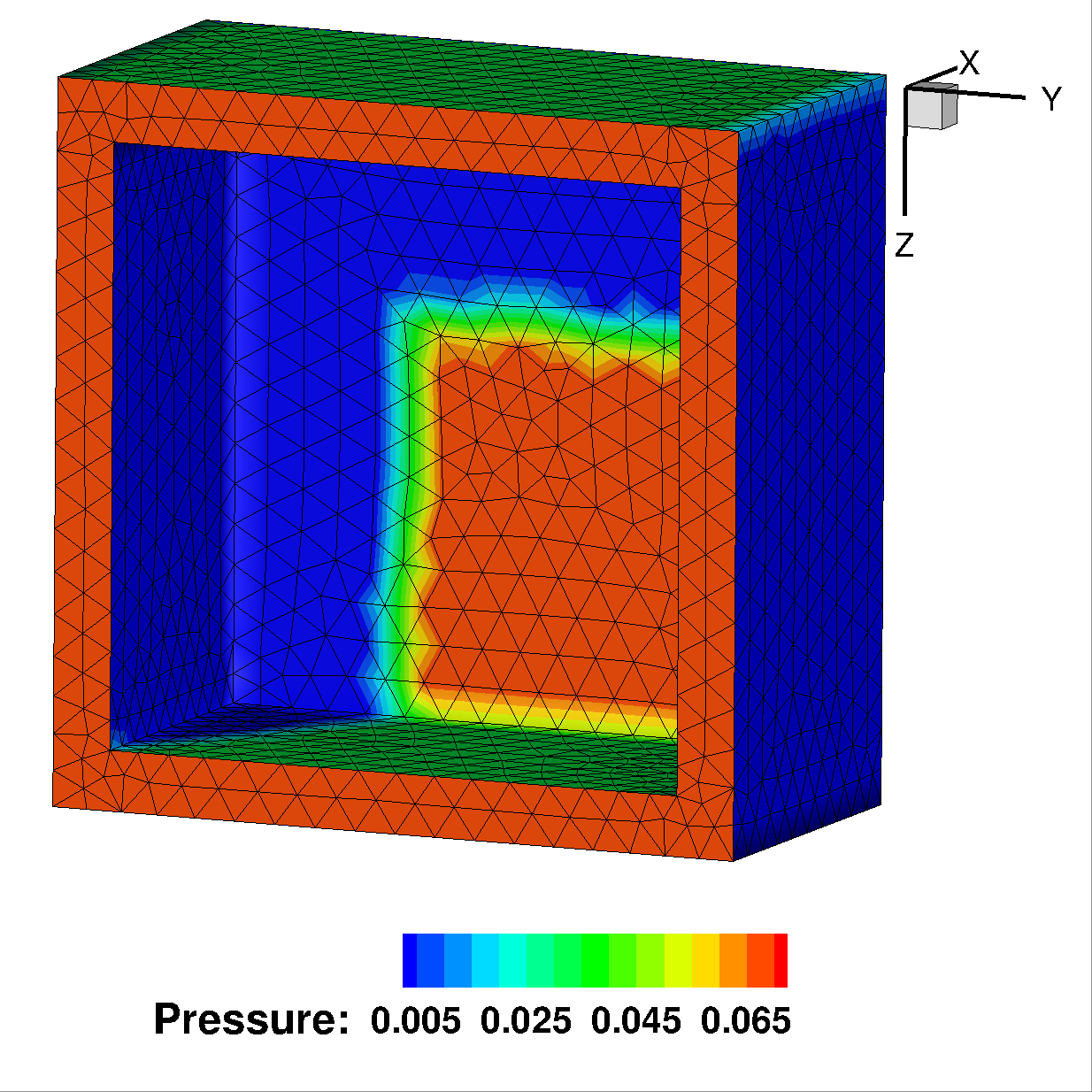}

{\small (a)}
\end{minipage}
\hfill
\begin{minipage}[t]{0.48\textwidth}
\centering
\includegraphics[width=\linewidth]{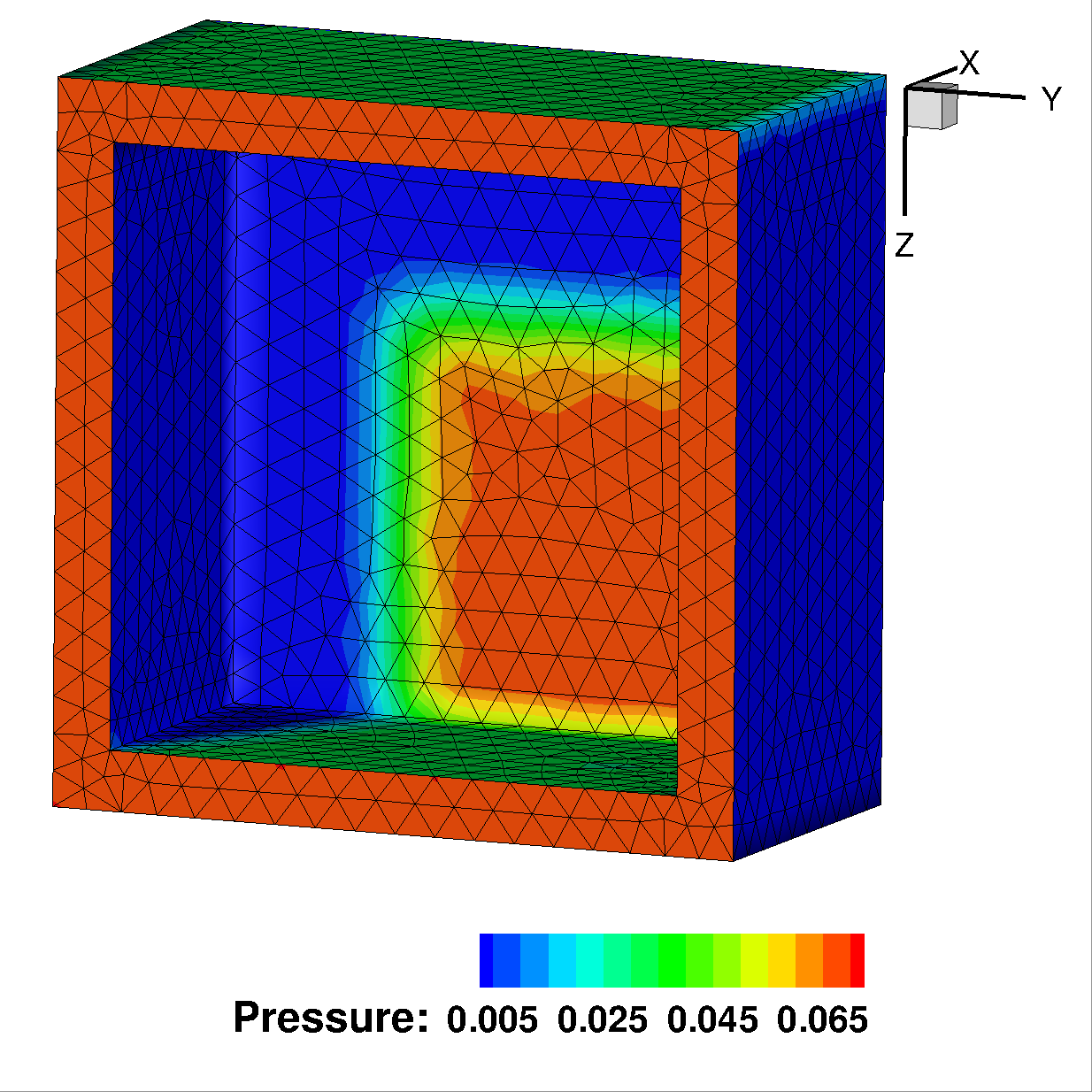}

{\small (b)}
\end{minipage}

\caption{Surface pressure for the open box at $(\mathrm{pitch},\mathrm{yaw},\mathrm{roll})=(30^\circ,30^\circ,0^\circ)$ from (a) GK and (b) MC. The cavity rim and walls shield parts of the interior.}
\label{fig:openbox_pressure}
\end{figure*}
In the O3D pressure maps (Fig.~\ref{fig:openbox_pressure}), the directly exposed cavity floor and portions of the interior walls carry the largest pressures. The upstream walls and outer frame shield nearby regions. Both solvers locate the main pressure transition along the cavity walls.

\subsection{CLL Flat-Plate Benchmark with SPARTA DSMC}
\label{sec:cll_dsmc_validation}

The inclined-plate benchmark of Woronowicz and Rault~\cite{WoronowiczRault1994} sets the conditions for an independent comparison of HACS CLL-exact, HACS MC, and SPARTA DSMC~\cite{Plimpton2019}. The freestream speed ratio is $S_\infty=\beta_s|\mathbf U_\infty|=10$, and $T_\infty/T_w=1$. The plate inclination $\theta$ is measured from the plate surface, so $90^\circ$ denotes normal incidence.
\begin{figure*}[htbp]
\centering

\begin{minipage}[t]{0.48\textwidth}
\centering
\includegraphics[width=\linewidth]{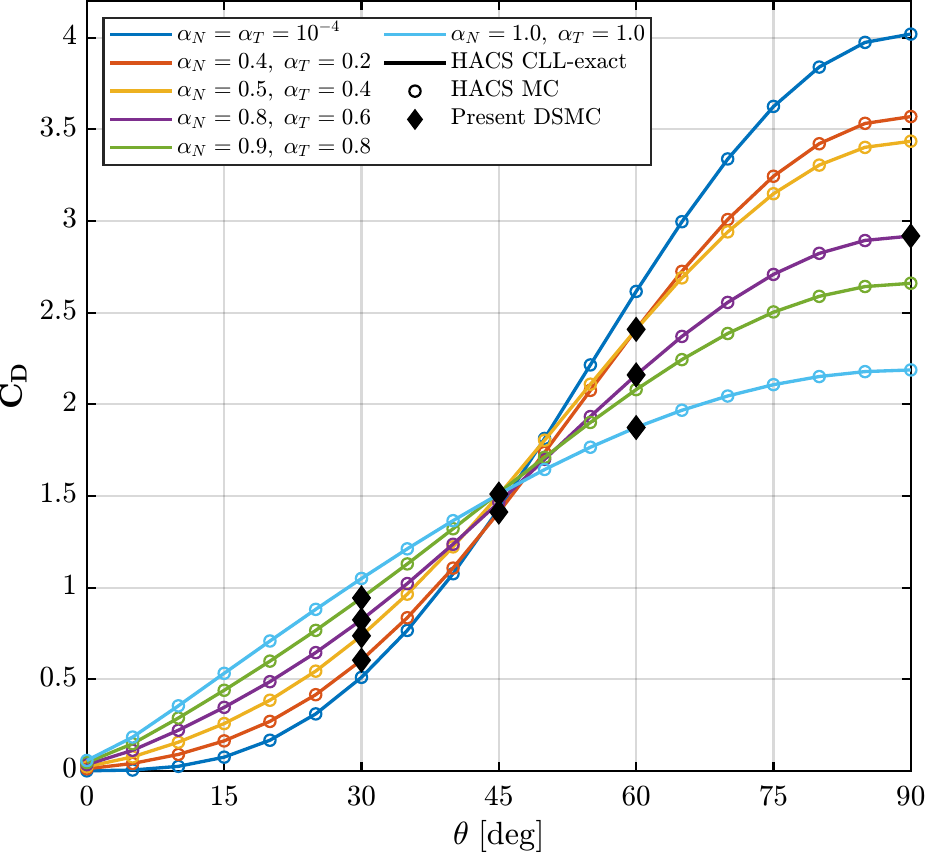}

{\small (a)}
\end{minipage}
\hfill
\begin{minipage}[t]{0.48\textwidth}
\centering
\includegraphics[width=\linewidth]{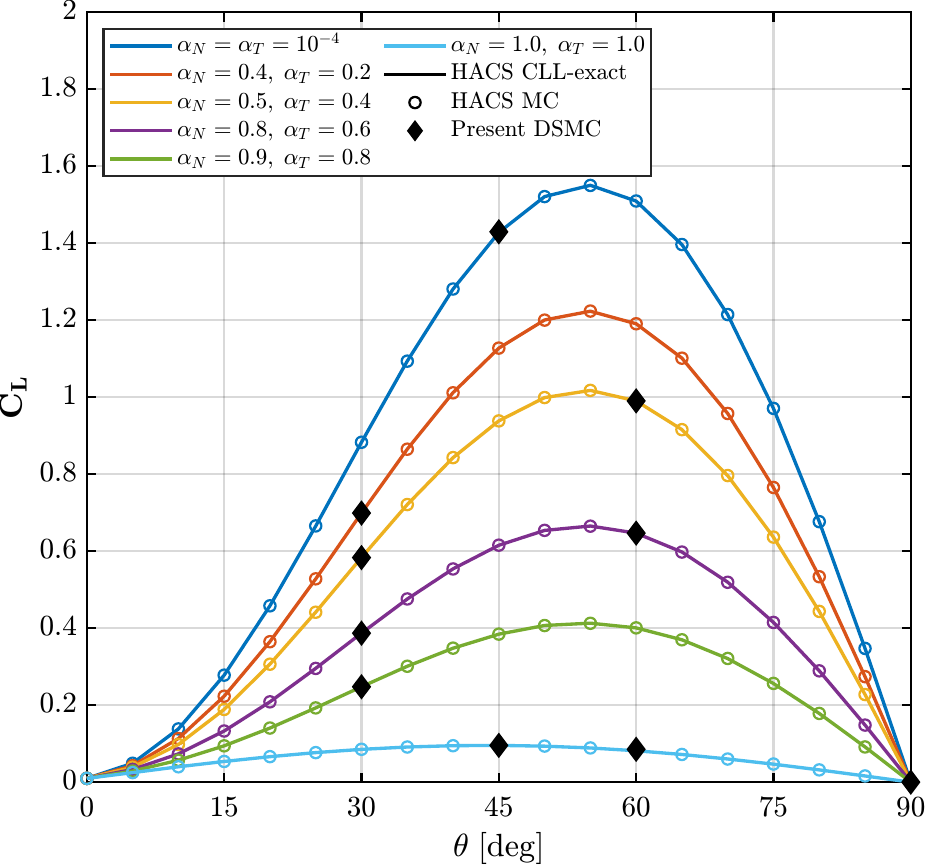}

{\small (b)}
\end{minipage}

\caption{CLL flat-plate benchmark at $S_\infty=10$ and $T_\infty/T_w=1$ using the conditions of Woronowicz and Rault~\cite{WoronowiczRault1994}. Panels (a) and (b) show drag and lift coefficients. Lines denote HACS CLL-exact, open symbols denote HACS MC, and filled symbols denote the present SPARTA DSMC results.}
\label{fig:cll_external_validation}
\end{figure*}
The compared CLL pairs $(\alpha_N,\alpha_T)$ are $(10^{-4},10^{-4})$, $(0.4,0.2)$, $(0.5,0.4)$, $(0.8,0.6)$, $(0.9,0.8)$, and $(1.0,1.0)$. The HACS predictions cover $0^\circ\le\theta\le90^\circ$. SPARTA results are shown as individual points at the conditions listed in Table~\ref{tab:cll_dsmc_validation}.

\begin{table*}[htbp]
\centering
\caption{Comparison of HACS CLL-exact and particle-sampling Monte Carlo predictions with the present SPARTA DSMC calculations for selected flat-plate CLL benchmark conditions. Values in parentheses denote the relative error with respect to the DSMC result.}
\label{tab:cll_dsmc_validation}

\footnotesize
\setlength{\tabcolsep}{3pt}

\begin{ruledtabular}
\begin{tabular}{ccccccc}
\multicolumn{1}{c}{CLL parameters} &
\multicolumn{2}{c}{DSMC} &
\multicolumn{2}{c}{HACS CLL-exact} &
\multicolumn{2}{c}{HACS MC-CLL} \\
\hline
$\alpha_N$, $\alpha_T$, $\theta$ &
$C_L$ &
$C_D$ &
$C_L$ (Err.\%) &
$C_D$ (Err.\%) &
$C_L$ (Err.\%) &
$C_D$ (Err.\%) \\
\hline
$10^{-4}$, $10^{-4}$, 45 &
1.43115 & 1.41903 &
1.42825 (0.203) &
1.42839 (0.660) &
1.42805 (0.217) &
1.42819 (0.645) \\
0.4, 0.2, 30 &
0.70171 & 0.59875 &
0.69943 (0.325) &
0.60382 (0.847) &
0.69930 (0.344) &
0.60375 (0.835) \\
0.5, 0.4, 30 &
0.58212 & 0.74151 &
0.58386 (0.298) &
0.73709 (0.596) &
0.58361 (0.257) &
0.73687 (0.626) \\
0.5, 0.4, 60 &
0.99548 & 2.42938 &
0.99064 (0.487) &
2.40865 (0.853) &
0.99065 (0.486) &
2.40886 (0.845) \\
0.8, 0.6, 30 &
0.38618 & 0.81891 &
0.38733 (0.300) &
0.82363 (0.576) &
0.38716 (0.255) &
0.82355 (0.567) \\
0.8, 0.6, 60 &
0.64650 & 2.17888 &
0.64715 (0.101) &
2.16013 (0.861) &
0.64720 (0.108) &
2.16009 (0.862) \\
0.8, 0.6, 90 &
0.00023 & 2.93296 &
0.00000$^{a}$ &
2.91789 (0.514) &
0.00023$^{a}$ &
2.91845 (0.495) \\
0.9, 0.8, 30 &
0.24770 & 0.95144 &
0.24808 (0.154) &
0.94323 (0.863) &
0.24805 (0.141) &
0.94311 (0.876) \\
1.0, 1.0, 60 &
0.08229 & 1.86918 &
0.08175 (0.660) &
1.87365 (0.239) &
0.08173 (0.681) &
1.87333 (0.222) \\
1.0, 1.0, 45 &
0.09665 & 1.49893 &
0.09569 (0.990) &
1.50991 (0.733) &
0.09574 (0.938) &
1.50995 (0.736) \\
\end{tabular}
\end{ruledtabular}

\vspace{1mm}

{\footnotesize
$^{a}$The theoretical lift vanishes by symmetry at $\theta=90^\circ$. The absolute lift coefficients are listed for this condition.
}

\end{table*}

The two HACS solutions are nearly coincident in Fig.~\ref{fig:cll_external_validation}. Across the ten DSMC comparisons, the mean relative differences in $C_D$ are $0.674\%$ for CLL-exact and $0.671\%$ for MC. Their maximum differences are $0.863\%$ and $0.876\%$.

For $C_L$, omitting the normal-incidence case where the theoretical lift vanishes, the mean relative differences are $0.391\%$ for CLL-exact and $0.397\%$ for MC. Their maxima are $0.990\%$ and $0.938\%$. Both solvers recover the shift in the magnitude and angular position of the lift maximum as accommodation changes.

At normal incidence for $(\alpha_N,\alpha_T)=(0.8,0.6)$, the analytical lift is zero by symmetry. SPARTA and HACS MC each give a small residual of about $2.3\times10^{-4}$ at the reported precision. The absolute coefficient difference is the relevant measure at this symmetry point.

\subsection{Verification Summary}
\label{sec:verification_summary}

The flat-plate and reflection-angle tests check flux weighting and scattering. The cube and open-box comparisons extend those checks to spatial integration and geometric exposure. The SPARTA flat-plate cases benchmark CLL lift and drag at matched numerical conditions. The spacecraft cases below examine how those calculations behave when visibility and accommodation vary across a more complex geometry.

\section{Case Studies}
\label{sec:case_studies}

The first configuration represents JAXA's SLATS spacecraft, with a compact bus and an extended solar array. The second includes several arrays and appendages that can shield one another as the attitude changes. The two geometries are used to examine how the direction of the incoming flow and the resulting changes in exposed area affect the forces, moments, and heating.

All spacecraft calculations use the CLL scattering model. Unless labeled conventional GK, the gas-kinetic results use TAMV. In TAMV, the contribution of each incident velocity depends on both its normal speed and its access to the surface [Eq.~\eqref{eq:tamv_rate}]. The pressure contours are used to locate changes in surface loading, and the integrated results show the combined effect of pressure and shear over the spacecraft.

\subsection{JAXA SLATS Model}
\label{sec:jaxa_case}

\subsubsection{Geometry and Computational Cases}
\label{sec:jaxa_geometry_cases}

\begin{figure}[htbp]
    \centering
    \includegraphics[width=0.82\textwidth]{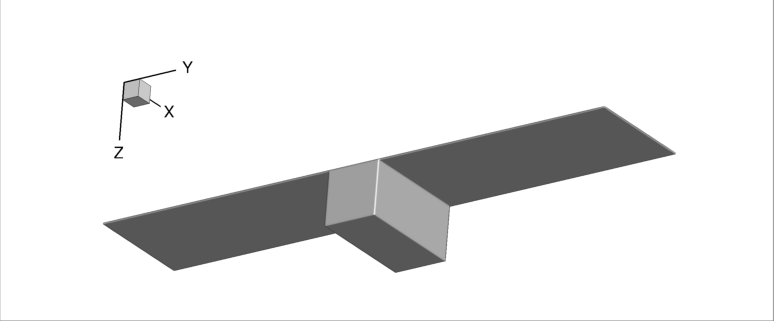}
    \caption{Triangulated JAXA SLATS geometry and body-fixed coordinate system used in the present rarefied-flow calculations.}
    \label{fig:jaxa_geometry}
\end{figure}

The simplified SLATS model consists of a compact body and an extended solar panel (Fig.~\ref{fig:jaxa_geometry}). The body axes are axial ($x$), spanwise ($y$), and normal ($z$). Pitch about the $y$-axis varies the freestream direction in the $x$--$z$ plane. Yaw about the $z$-axis varies it in the $x$--$y$ plane.

The SLATS cases in Table~\ref{tab:jaxa_cases} cover pitch and yaw at several altitudes and a separate accommodation sweep at 150~km. The baseline CLL parameters are $(\alpha_N,\alpha_T,\alpha_E)=(0.8,0.6,1.0)$.

\begin{table*}[htbp]
\centering
\caption{Definition of the JAXA SLATS case studies.}
\label{tab:jaxa_cases}
\small
\begin{ruledtabular}
\begin{tabular}{lllll}
Case & Altitude (km) & Attitude variation & $(\alpha_N,\alpha_T,\alpha_E)$ & Solvers \\
\hline
J1 & 120, 150, 170, 200, 250, 300 & Pitch: $-90^\circ$ to $90^\circ$ & $(0.8,0.6,1.0)$ & GK--TAMV, MC, DSMC$^{a}$ \\
J2 & 150, 200, 250 & Yaw: $-90^\circ$ to $90^\circ$ & $(0.8,0.6,1.0)$ & GK--TAMV, MC, DSMC$^{a}$ \\
J3 & 150 & Pitch: $-90^\circ$ to $90^\circ$ & C1--C4 & GK--TAMV, MC, DSMC$^{a}$ \\
\end{tabular}
\end{ruledtabular}
\vspace{1mm}
\begin{flushleft}
\footnotesize $^{a}$The present DSMC simulations were performed only at selected spacecraft attitudes.
\end{flushleft}
\end{table*}

Normal and tangential accommodation decrease across the four 150~km cases. The pairs $(\alpha_N,\alpha_T)$ are $(1.0,1.0)$ for C1, $(0.8,0.6)$ for C2, $(0.5,0.4)$ for C3, and $(0.2,0.2)$ for C4. The internal-energy accommodation is $\alpha_E=1$ in all four cases.

\subsubsection{Assessment of TAMV and Surface Pressure Distribution}
\label{sec:jaxa_tamv}

\begin{figure*}[htbp]
\centering
\subfloat[Pitch sweep]{%
  \includegraphics[width=0.48\textwidth]{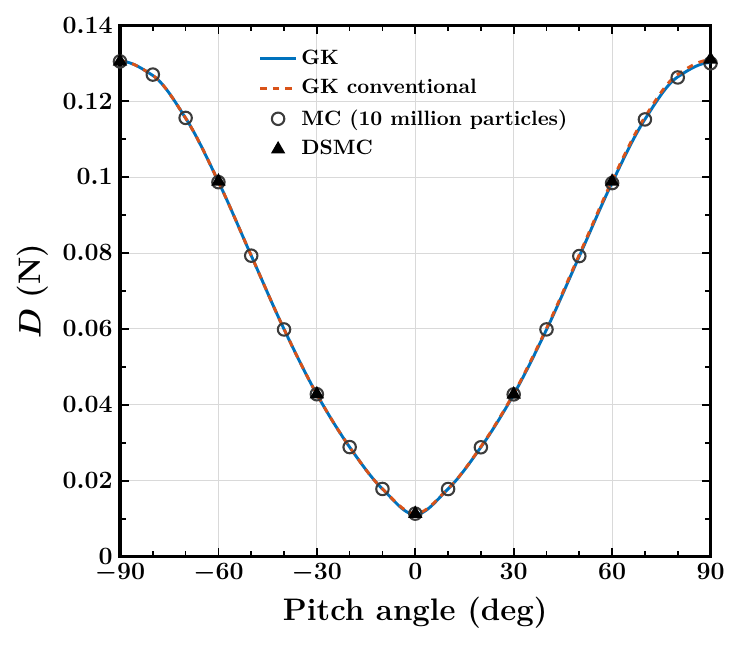}%
}
\hfill
\subfloat[Yaw sweep]{%
  \includegraphics[width=0.48\textwidth]{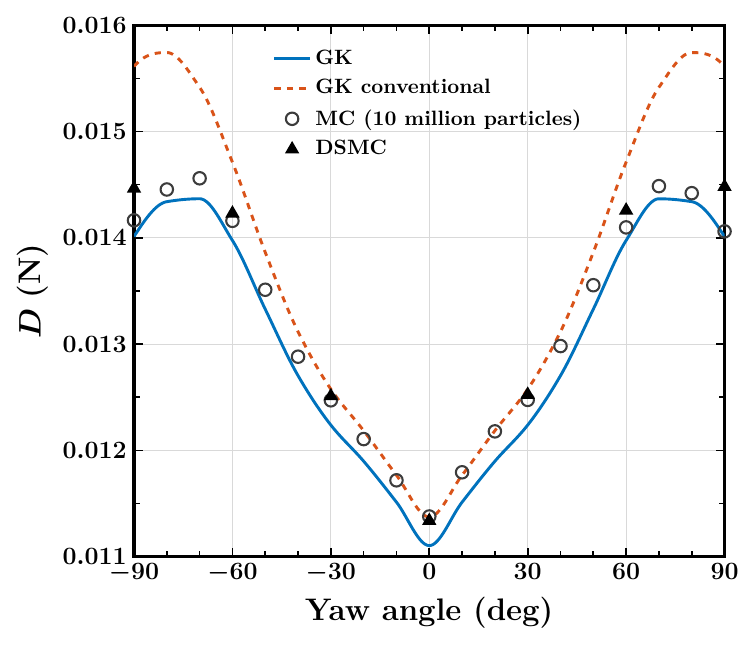}%
}
\caption{SLATS drag at 200~km during (a) the pitch sweep and (b) the yaw sweep. The solid curve labeled GK represents GK--TAMV. Open circles show MC results obtained with $10^7$ particles, and filled triangles show DSMC results at selected attitudes.}
\label{fig:jaxa_tamv_assessment}
\end{figure*}

Figure~\ref{fig:jaxa_tamv_assessment} compares the SLATS drag over the pitch and yaw sweeps at 200~km. During the pitch sweep [Fig.~\ref{fig:jaxa_tamv_assessment}(a)], rotation exposes the broad solar-array surface to a progressively larger normal component of the mean flow. The incident normal momentum flux increases because more molecules cross a unit surface area per unit time and each impact carries more normal momentum [Eqs.~\eqref{eq:incident_number_flux} and \eqref{eq:incident_pressure}]. Drag rises from about $0.011~\mathrm{N}$ at zero pitch to $0.13~\mathrm{N}$ at $\pm90^\circ$. The two GK curves nearly overlap and agree closely with MC and the available DSMC points. At $-60^\circ$ pitch, all four methods give about $0.10~\mathrm{N}$.

Yaw changes the incidence on the lateral faces of the bus in the $x$--$y$ plane, with the broad array remaining close to grazing incidence. The drag therefore varies over a smaller range [Fig.~\ref{fig:jaxa_tamv_assessment}(b)]. Near zero yaw the two GK predictions are close. At larger yaw magnitudes, conventional GK gives higher drag, and GK--TAMV approaches the particle results more closely. At $-60^\circ$ yaw, MC and DSMC both give approximately $0.0142~\mathrm{N}$. The greater separation between the GK curves in yaw is consistent with increased sensitivity to direction-dependent shielding when the mean flow is nearly tangent to part of the spacecraft.

\begin{figure*}[htbp]
\centering
\subfloat[GK--TAMV]{%
  \includegraphics[width=0.48\textwidth]{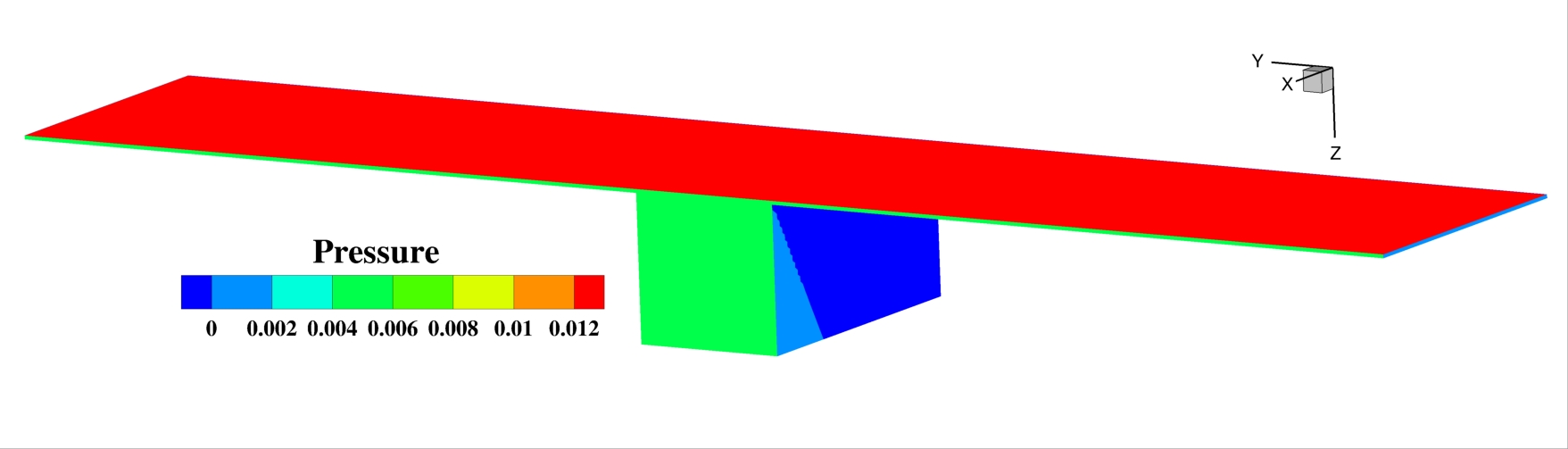}%
}
\hfill
\subfloat[Conventional GK]{%
  \includegraphics[width=0.48\textwidth]{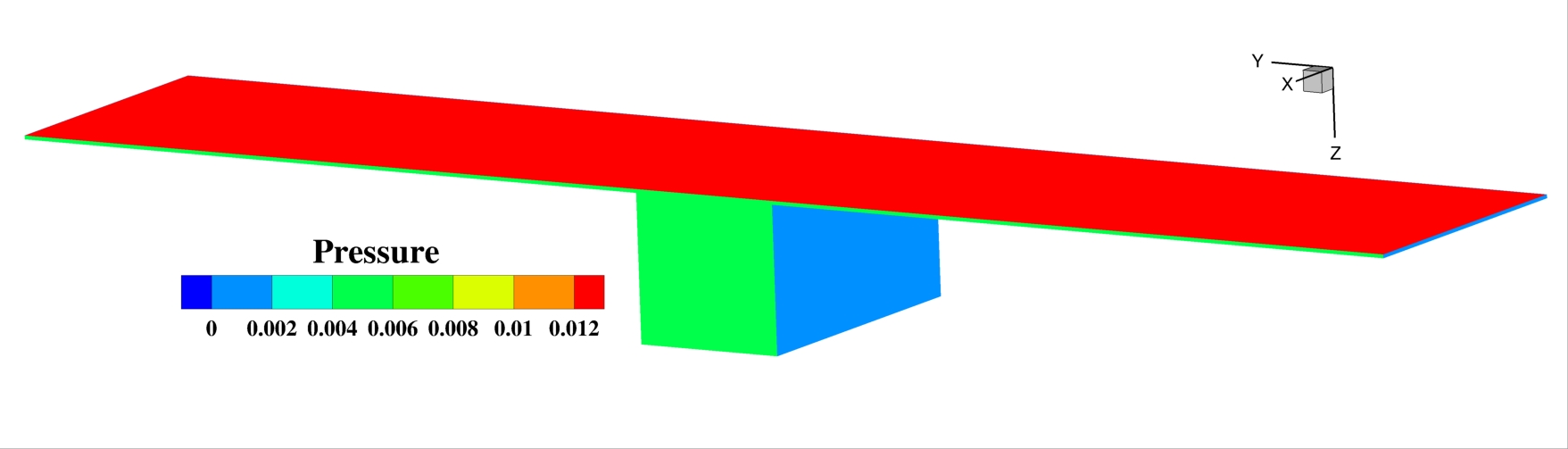}%
}

\vspace{2mm}

\subfloat[MC]{%
  \includegraphics[width=0.48\textwidth]{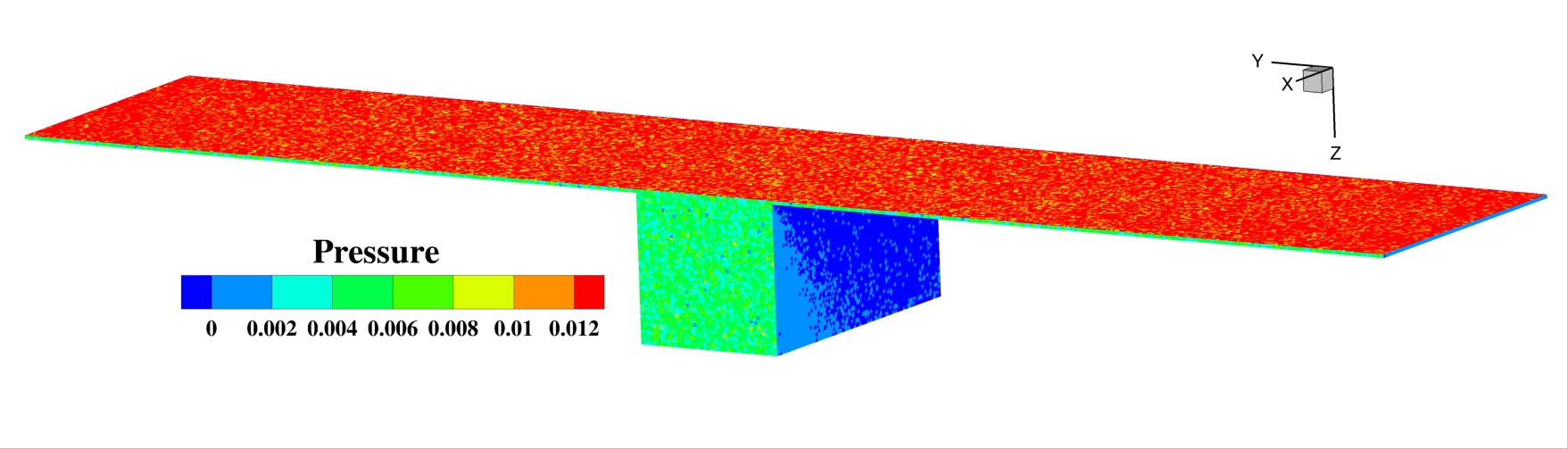}%
}
\hfill
\subfloat[DSMC]{%
  \includegraphics[width=0.48\textwidth]{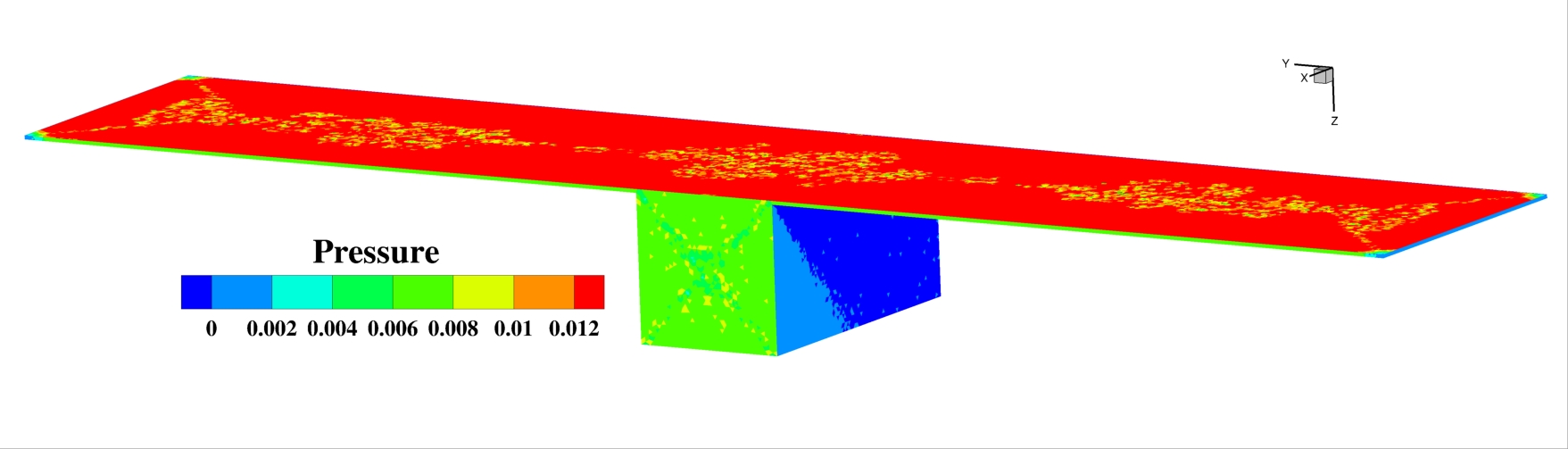}%
}
\caption{SLATS surface pressure at 200~km and a pitch angle of $-60^\circ$ from (a) GK--TAMV, (b) conventional GK, (c) MC, and (d) DSMC.}
\label{fig:jaxa_pressure_comparison}
\end{figure*}

Figure~\ref{fig:jaxa_pressure_comparison} shows the pressure distributions at $-60^\circ$ pitch. The solar array receives the highest pressure in all four solutions, consistent with the large inward component of the incident velocity on its exposed face. On the more lightly loaded bus side, conventional GK gives a nearly uniform pressure. GK--TAMV resolves a lower-pressure region that is also visible in the MC and DSMC contours. The local change in pressure on the bus side is confined to a smaller, less heavily loaded area. Its contribution to drag is further weighted by the projection of the surface normal onto the freestream direction, which helps explain why the total drag remains nearly the same.

\begin{figure*}[htbp]
\centering
\subfloat[GK--TAMV]{%
  \includegraphics[width=0.48\textwidth]{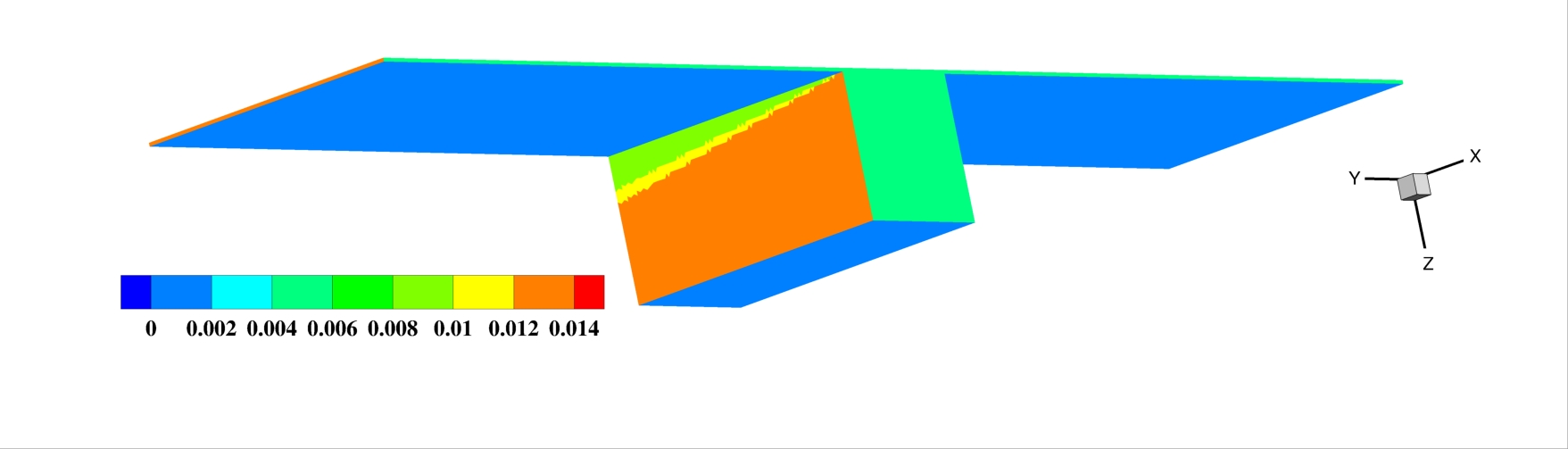}%
}
\hfill
\subfloat[Conventional GK]{%
  \includegraphics[width=0.48\textwidth]{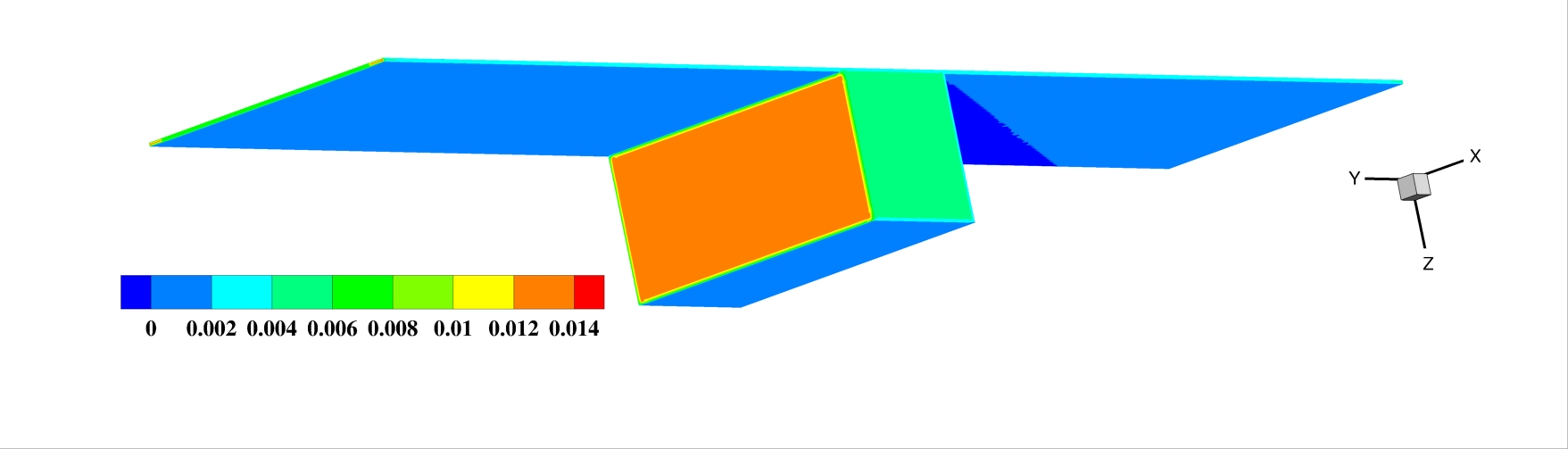}%
}

\vspace{2mm}

\subfloat[MC]{%
  \includegraphics[width=0.48\textwidth]{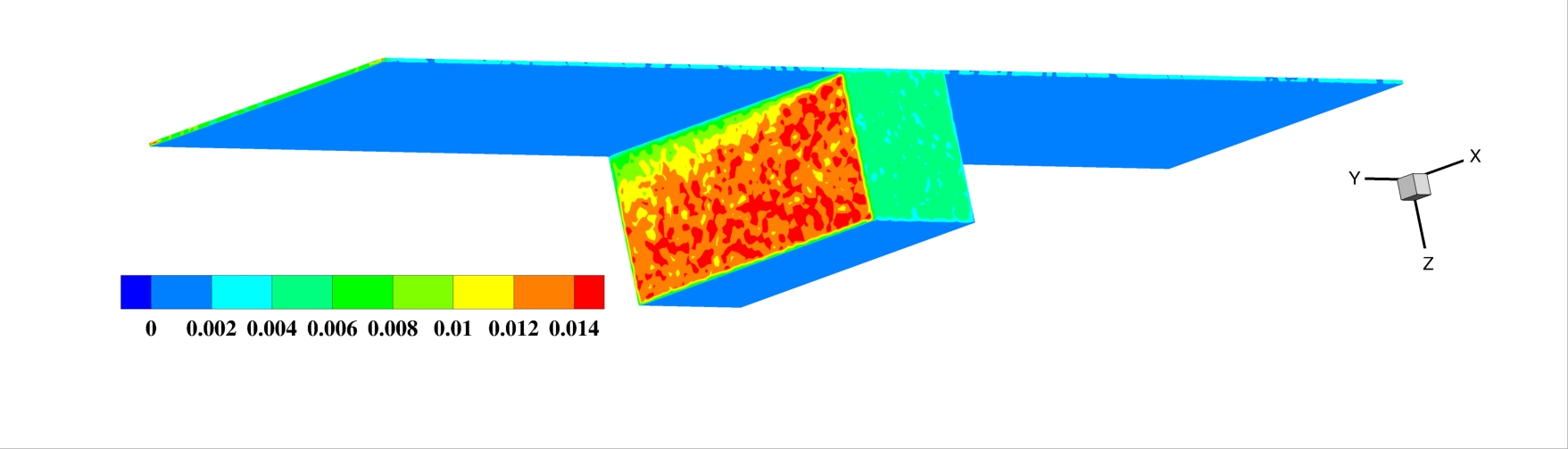}%
}
\hfill
\subfloat[DSMC]{%
  \includegraphics[width=0.48\textwidth]{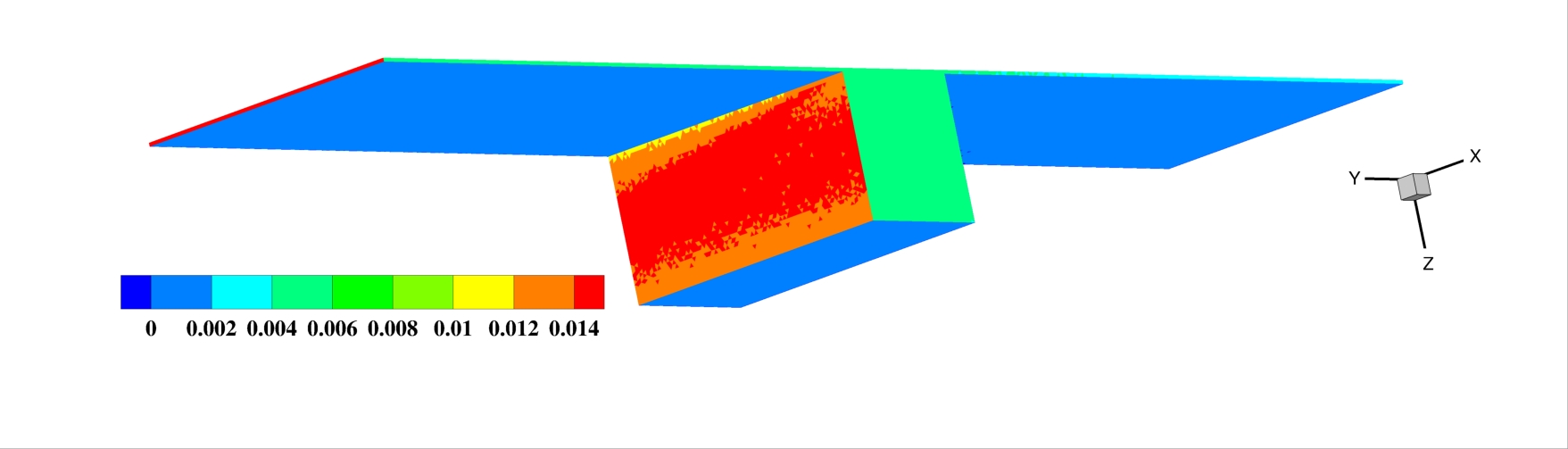}%
}
\caption{SLATS surface pressure at 200~km and a yaw angle of $-60^\circ$ from (a) GK--TAMV, (b) conventional GK, (c) MC, and (d) DSMC.}
\label{fig:jaxa_pressure_comparison_yaw}
\end{figure*}

At $-60^\circ$ yaw (Fig.~\ref{fig:jaxa_pressure_comparison_yaw}), the exposed bus face experiences stronger normal incidence and carries the highest pressure. Most of the array remains lightly loaded. Conventional GK predicts a triangular region of near-zero pressure on the array next to the bus because the bus blocks trajectories aligned with the mean flow. In TAMV, some molecules can reach this region with a thermal velocity component transverse to the mean flow. Their contribution is weighted by the inward normal speed in Eq.~\eqref{eq:tamv_rate}, giving a finite pressure across the region. The MC and DSMC maps show a similar spatial pattern there, and DSMC gives higher pressure over much of the exposed bus face than either GK result. The drag difference depends on the sum of pressure and shear projected along the freestream over the entire surface.

\subsubsection{Combined Altitude and Pitch-Angle Effects}
\label{sec:jaxa_pitch}

\begin{figure*}[htbp]
\centering
\includegraphics[width=0.49\textwidth]{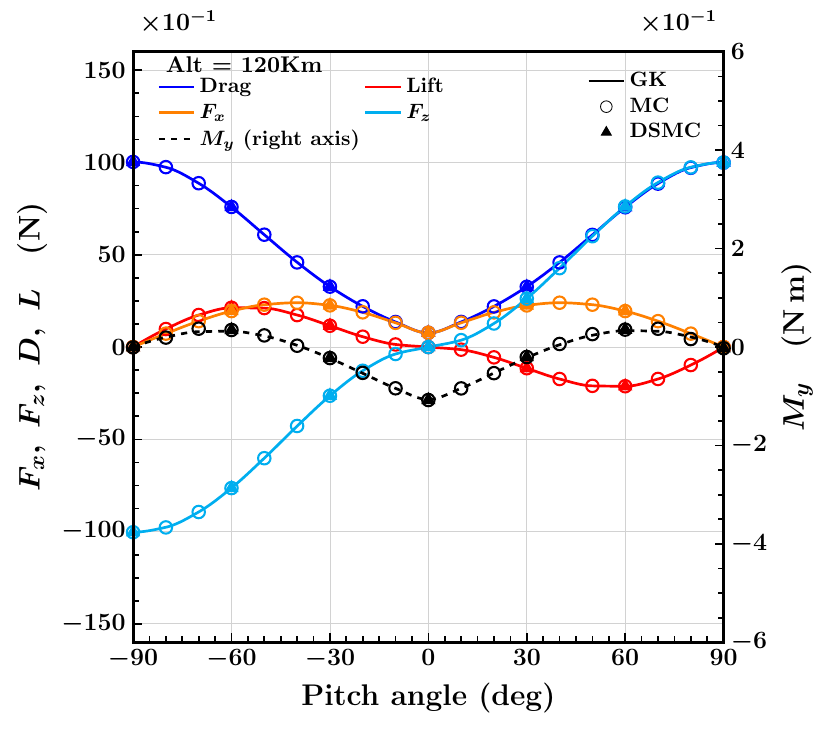}\hfill
\includegraphics[width=0.49\textwidth]{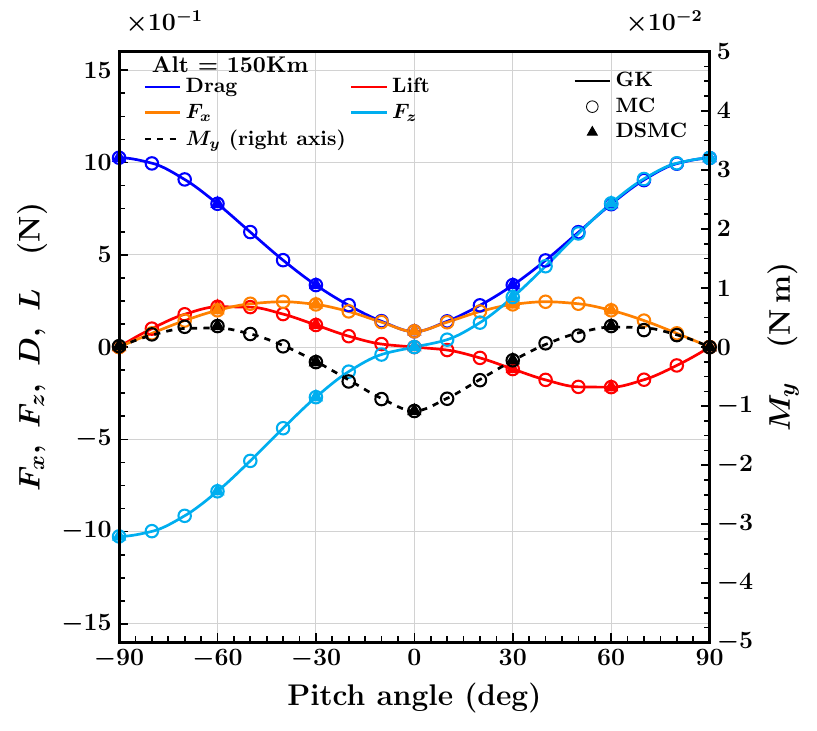}\\[3pt]
\includegraphics[width=0.49\textwidth]{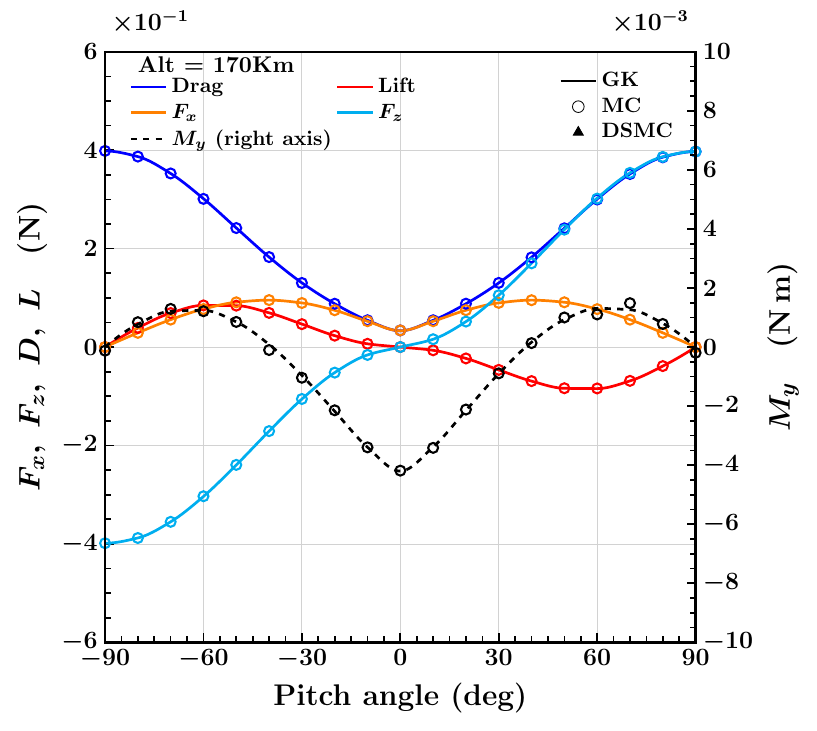}\hfill
\includegraphics[width=0.49\textwidth]{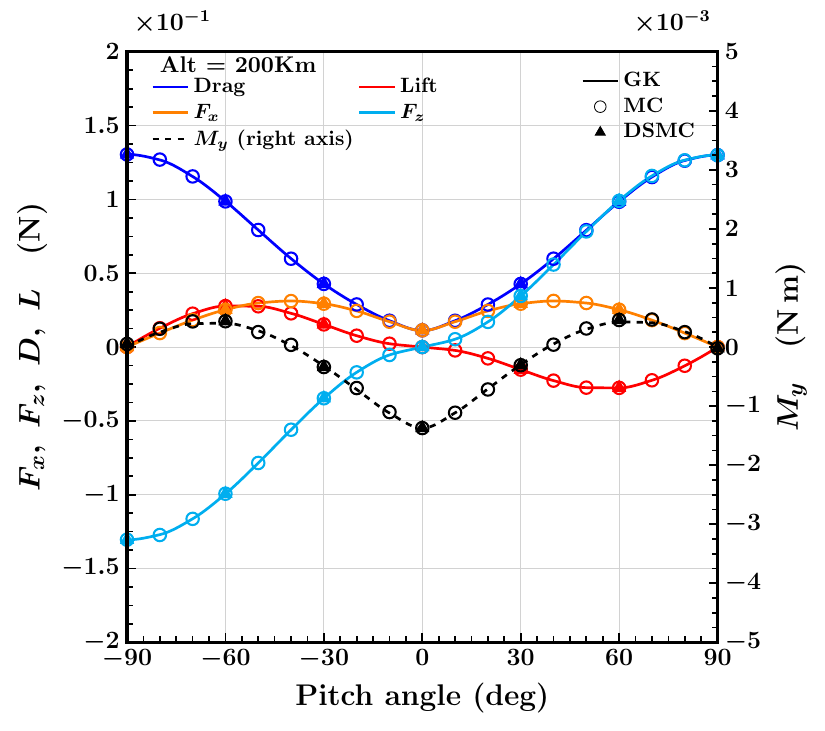}\\[3pt]
\includegraphics[width=0.49\textwidth]{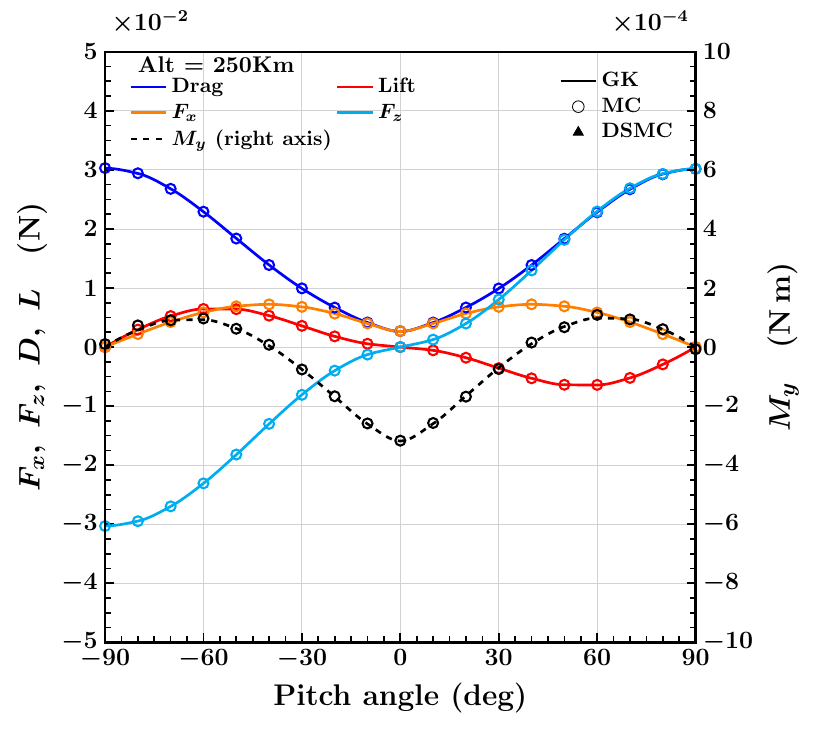}\hfill
\includegraphics[width=0.49\textwidth]{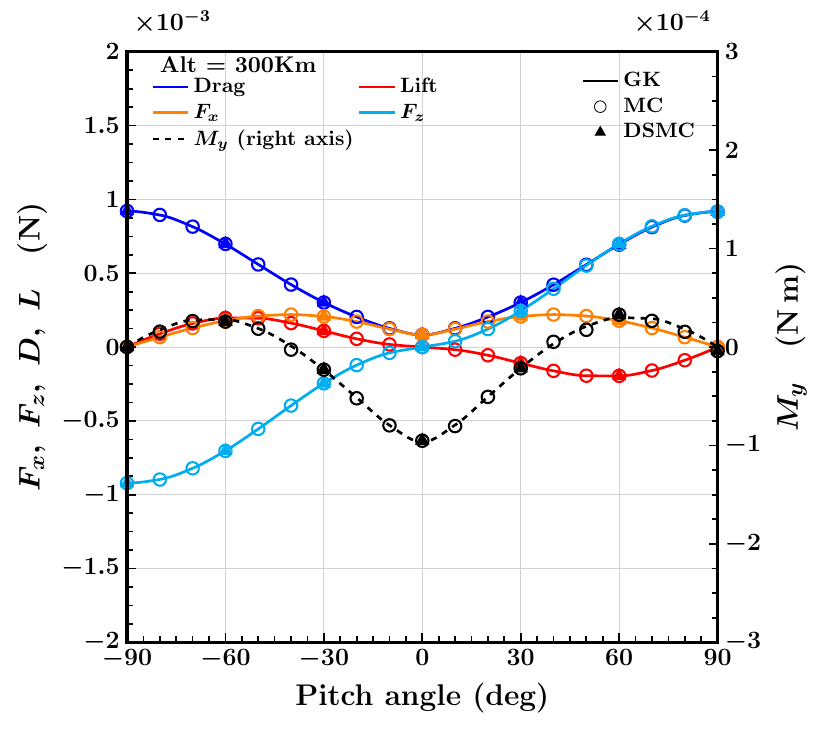}
\caption{Pitch-angle dependence of the SLATS aerodynamic loads at (a) 120, (b) 150, (c) 170, (d) 200, (e) 250, and (f) 300~km. Drag, lift, $F_x$, $F_z$, and $M_y$ are compared using GK--TAMV, MC, and available DSMC results.}
\label{fig:jaxa_pitch_forces}
\end{figure*}

The pitch sweeps from 120 to 300~km (Fig.~\ref{fig:jaxa_pitch_forces}) show a decrease in the dimensional loads with increasing altitude. At a given pitch angle, the reduction in freestream species densities lowers the number of impacts and hence the momentum transferred per unit time. The angular trends are similar at the six altitudes because pitching still changes the exposed area and the direction of the force on the same surfaces.

The body-normal force $F_z$ and lift reverse sign as pitching reverses the normal component of the mean flow over the broad array. The axial force remains a major component of the integrated load. The pitching moment $M_y$ also depends on where the force acts. In Eq.~\eqref{eq:force_moment_total}, the panel contributions to $M_y$ are $r_zF_x-r_xF_z$, so changes in the distribution of axial and normal loading can modify the moment independently of the drag trend. GK--TAMV and MC follow similar force and moment curves at the six altitudes, and the DSMC points shown at selected conditions follow the same general trends.

\begin{figure}[htbp]
    \centering
    \includegraphics[width=0.72\textwidth]{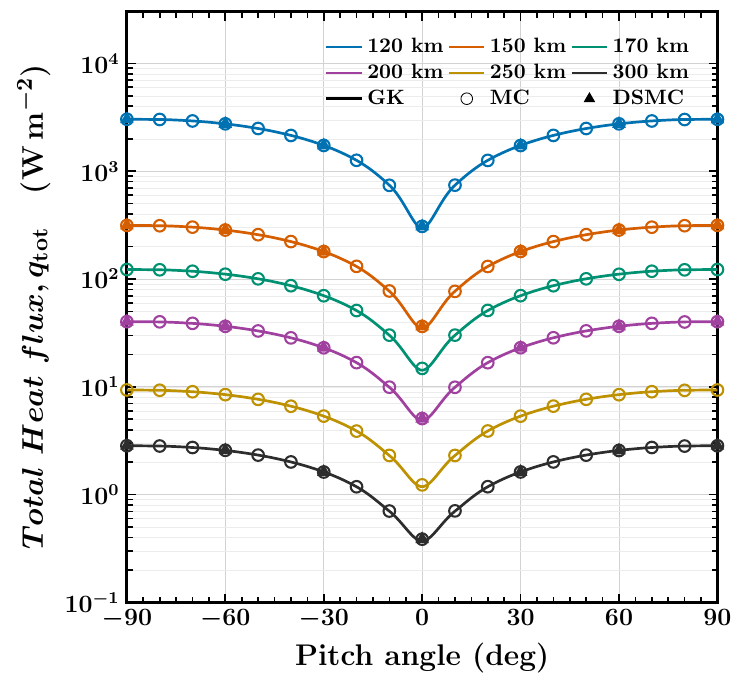}
    \caption{SLATS heat flux ($\mathrm{W\,m^{-2}}$) versus pitch at 120, 150, 170, 200, 250, and 300~km. The plot includes DSMC results at selected altitudes.}
    \label{fig:jaxa_pitch_heat}
\end{figure}

The heating results in Fig.~\ref{fig:jaxa_pitch_heat} show a steep decrease with altitude and a minimum near the nominal pitch attitude. The incident translational energy carried by the mean flow scales with $\rho_\infty|\mathbf U_\infty|^3/2$ per unit area normal to that flow. The energy reaching the spacecraft depends on the exposed area and incidence, and the net heating follows after subtracting the energy carried away by reflected molecules. The lower density at higher altitudes reduces the incident energy flux, and increasing the pitch magnitude exposes more of the array and increases the net heating.

\subsubsection{Combined Altitude and Yaw-Angle Effects}
\label{sec:jaxa_yaw}

\begin{figure*}[htbp]
    \centering
    \subfloat[150~km]{\includegraphics[width=0.48\textwidth]{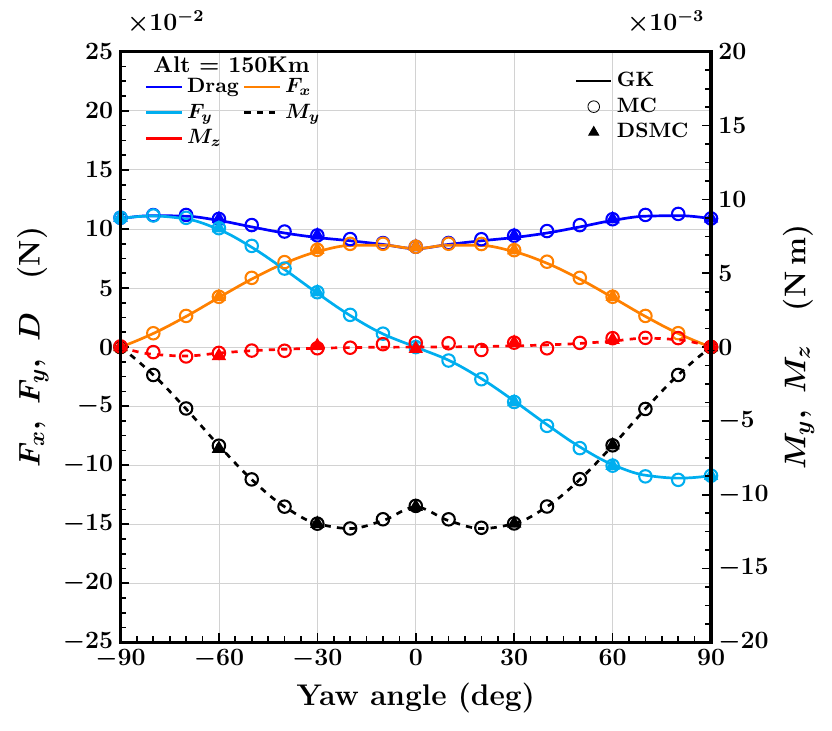}}
    \hfill
    \subfloat[200~km]{\includegraphics[width=0.48\textwidth]{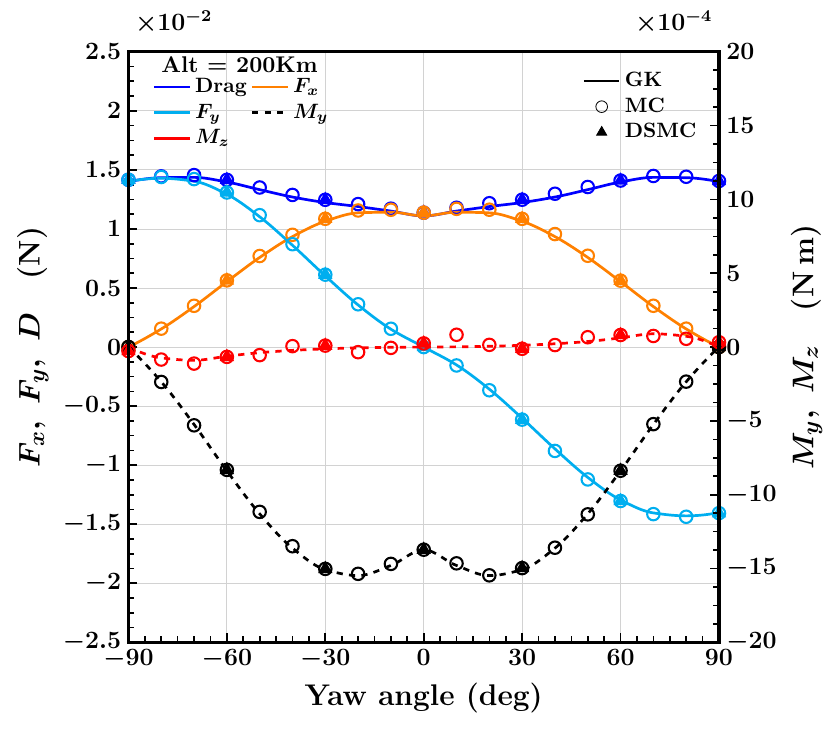}}
    \\[3pt]
    \subfloat[250~km]{\includegraphics[width=0.48\textwidth]{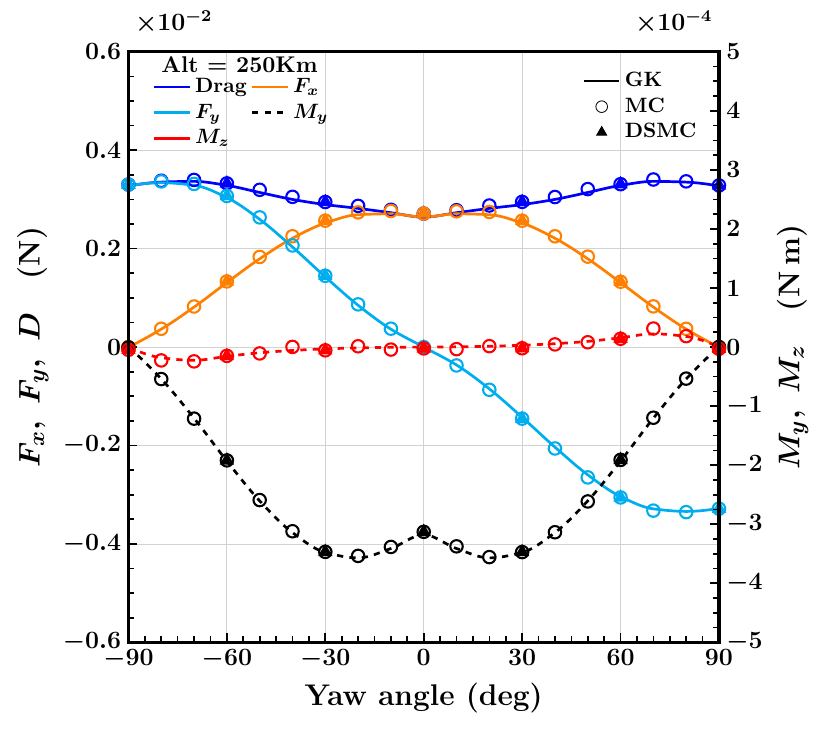}}
    \hfill
    \subfloat[Total heat flux]{\includegraphics[width=0.45\textwidth]{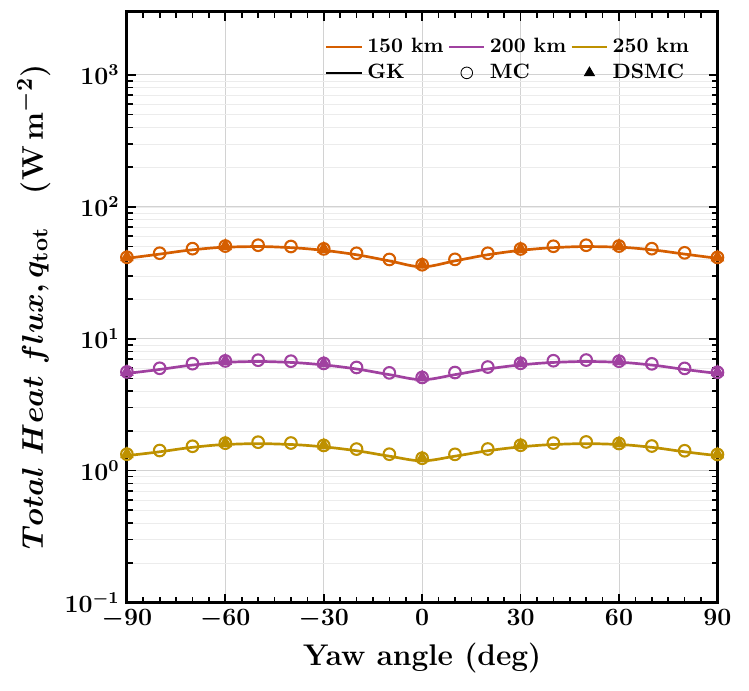}}
    \caption{Yaw-angle dependence of the aerodynamic and aerothermal response of the JAXA SLATS configuration at 150, 200, and 250~km.}
    \label{fig:jaxa_yaw}
\end{figure*}

The yaw sweeps at 150, 200, and 250~km give an approximately antisymmetric $F_y$ response (Fig.~\ref{fig:jaxa_yaw}), as reversing the yaw angle switches the direction of the lateral pressure loading. Drag and the axial force remain approximately symmetric about zero yaw because their dominant contributions have the same sign when the flow approaches from the opposite lateral direction. The yawing moment $M_z$ contains both $r_xF_y$ and $-r_yF_x$, so axial loads at opposite spanwise positions can partly cancel. An axial load acting above or below the moment origin also contributes to $M_y$ through $r_zF_x$.

The dimensional loads fall as the number of molecular impacts decreases with altitude, and GK--TAMV closely follows MC over the three yaw sweeps.

\subsubsection{Accommodation-Coefficient Effects at 150~km}
\label{sec:jaxa_accommodation}

\begin{figure*}[htbp]
\centering
\includegraphics[width=0.32\textwidth]{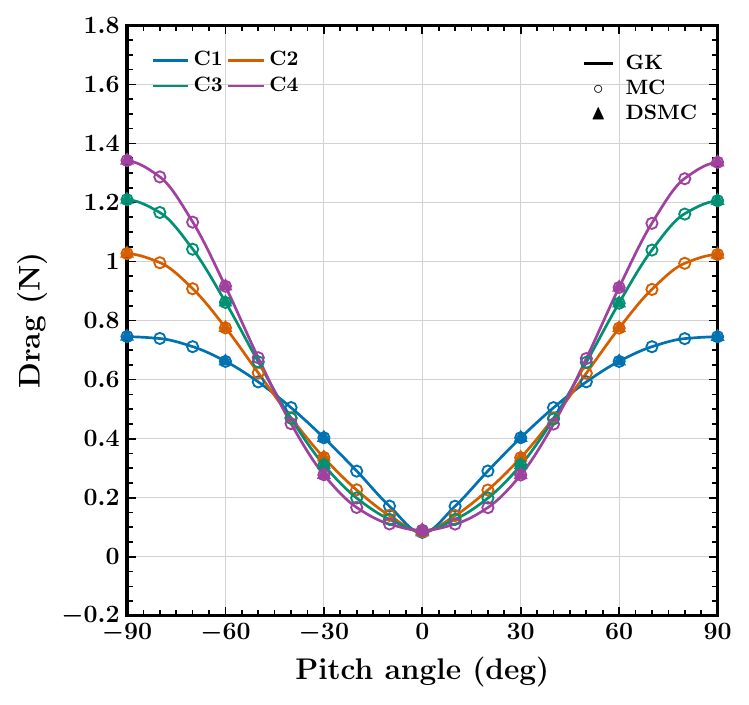}\hfill
\includegraphics[width=0.32\textwidth]{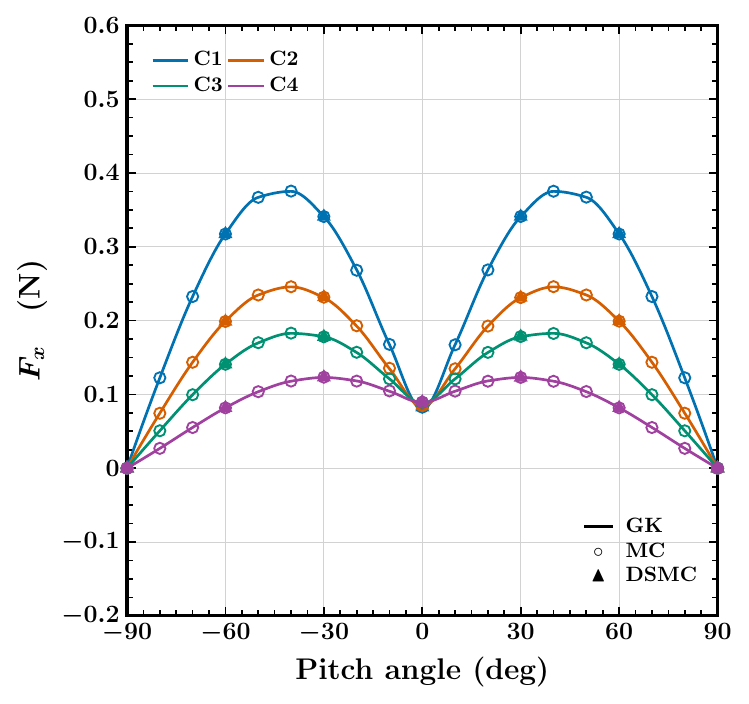}\hfill
\includegraphics[width=0.32\textwidth]{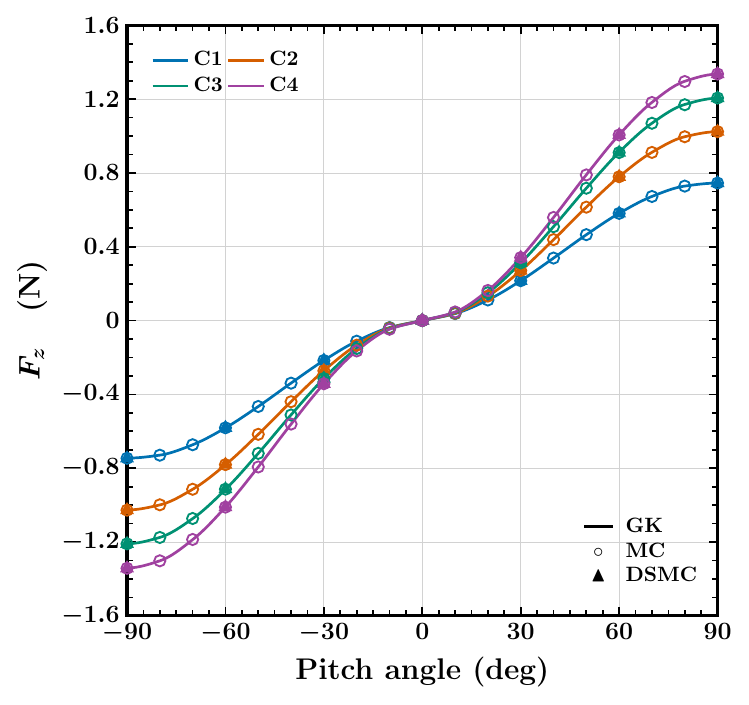}\\[3pt]
\includegraphics[width=0.32\textwidth]{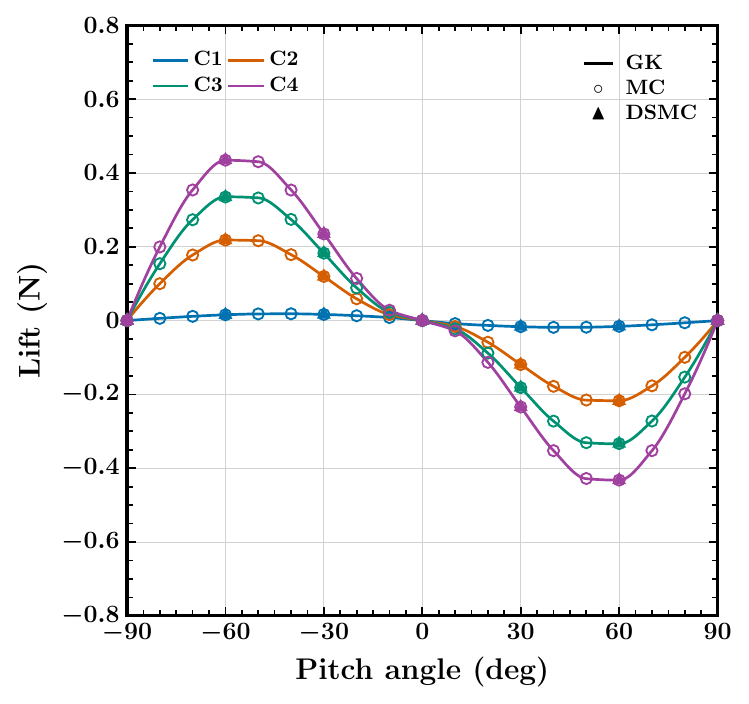}\hfill
\includegraphics[width=0.32\textwidth]{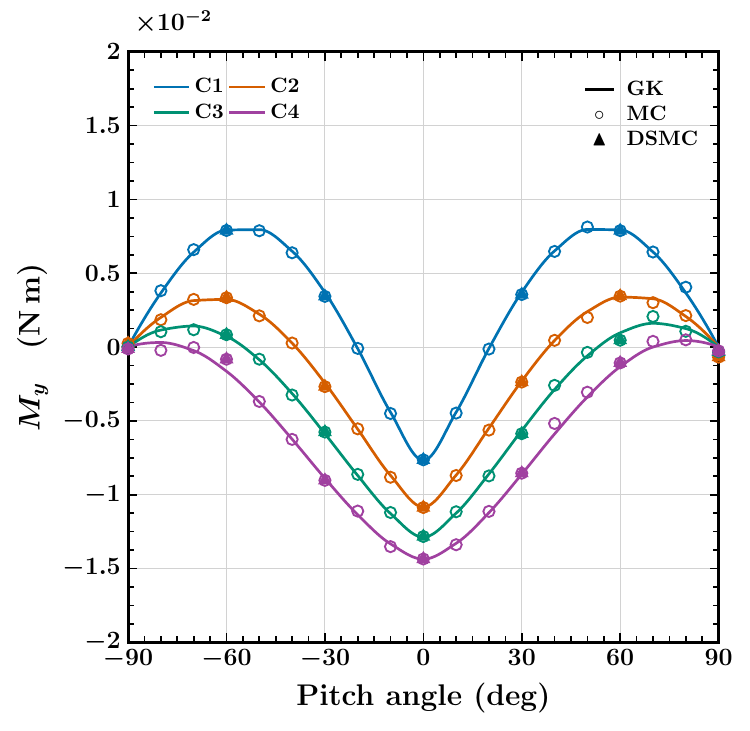}\hfill
\includegraphics[width=0.32\textwidth]{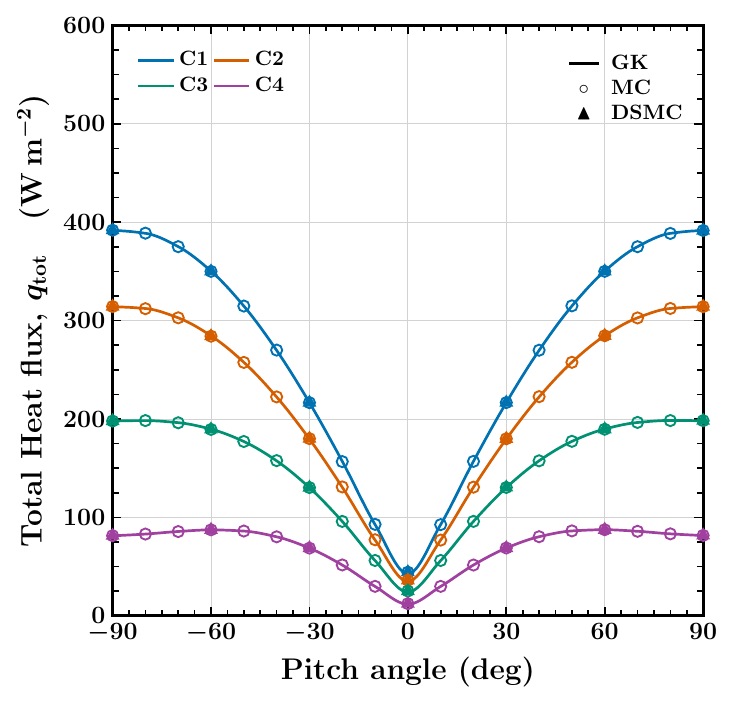}
\caption{Influence of CLL accommodation coefficients on the JAXA SLATS response at 150~km: (a) drag, (b) $F_x$, (c) $F_z$, (d) lift, (e) pitching moment $M_y$, and (f) total heat flux.}
\label{fig:jaxa_accommodation}
\end{figure*}

The four CLL parameter sets in Fig.~\ref{fig:jaxa_accommodation} use the same incident flow at each pitch angle. The differences therefore come from the momentum and energy carried away by reflected molecules. The normal accommodation $\alpha_N$ changes the reflected normal-speed distribution and hence the normal pressure. The tangential accommodation $\alpha_T$ controls how much of the incident tangential momentum is transferred to the wall [Eq.~\eqref{eq:cll_shear}]. The load differences become more apparent at large pitch magnitudes, where the array receives a larger incident momentum flux.

The heating curves also separate at fixed $\alpha_E=1$. The normal and tangential parameters change the reflected translational energy [Eq.~\eqref{eq:cll_heat_trans}], which enters the net heat balance together with the internal-energy exchange. GK--TAMV and MC give similar trends for the four parameter sets.

\subsection{Complex Satellite}
\label{sec:complex_satellite}

\subsubsection{Geometry and Computational Cases}
\label{sec:complex_geometry_cases}

\begin{figure}[htbp]
    \centering
    \includegraphics[width=0.86\textwidth]{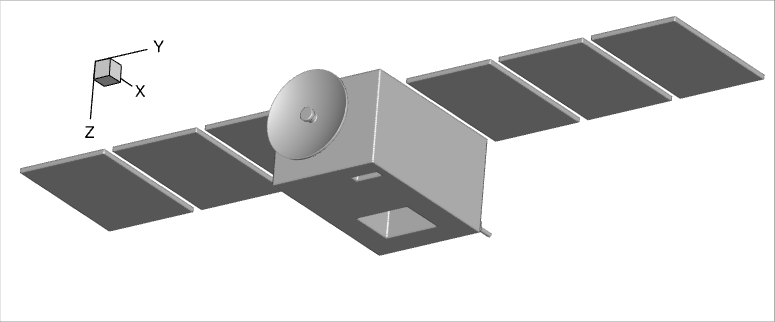}
    \caption{Complex satellite configuration and body-fixed coordinate system used in the present case study.}
    \label{fig:complex_geometry}
\end{figure}

The second spacecraft has a central bus, a multi-panel solar array, and several appendages. As the attitude changes, each component changes its projected area to the flow and can also shield surfaces farther downstream. The resulting changes in the location of the surface forces affect the moments through their lever arms about the moment origin. Figure~\ref{fig:complex_geometry} shows the body-fixed axes. Pitch and yaw use the same conventions as for SLATS.

Table~\ref{tab:complex_cases} lists the pitch sweep (CS1), yaw sweep (CS2), and accommodation sweep (CS3).

\begin{table*}[htbp]
\centering
\caption{Definition of the complex-satellite case studies.}
\label{tab:complex_cases}
\small
\begin{ruledtabular}
\begin{tabular}{lllll}
Case & Altitude (km) & Attitude variation & $(\alpha_N,\alpha_T,\alpha_E)$ & Solvers \\
\hline
CS1 & 120, 150, 180, 240, 300 & Pitch: $-90^\circ$ to $90^\circ$ & $(0.8,0.6,1.0)$ & GK--TAMV, MC \\
CS2 & 120, 180, 240 & Yaw: $-90^\circ$ to $90^\circ$ & $(0.8,0.6,1.0)$ & GK--TAMV, MC, DSMC$^{a}$ \\
CS3 & 150 & Pitch: $-90^\circ$ to $90^\circ$ & C1--C4 & GK--TAMV, MC \\
\end{tabular}
\end{ruledtabular}
{\footnotesize $^{a}$The DSMC pressure comparison shown for CS2 is at 180~km and $\psi=-60^\circ$.}\par
\end{table*}

\subsubsection{Combined Altitude and Pitch-Angle Effects}
\label{sec:complex_pitch}

\begin{figure*}[htbp]
    \centering
    \subfloat[120~km]{\includegraphics[width=0.48\textwidth]{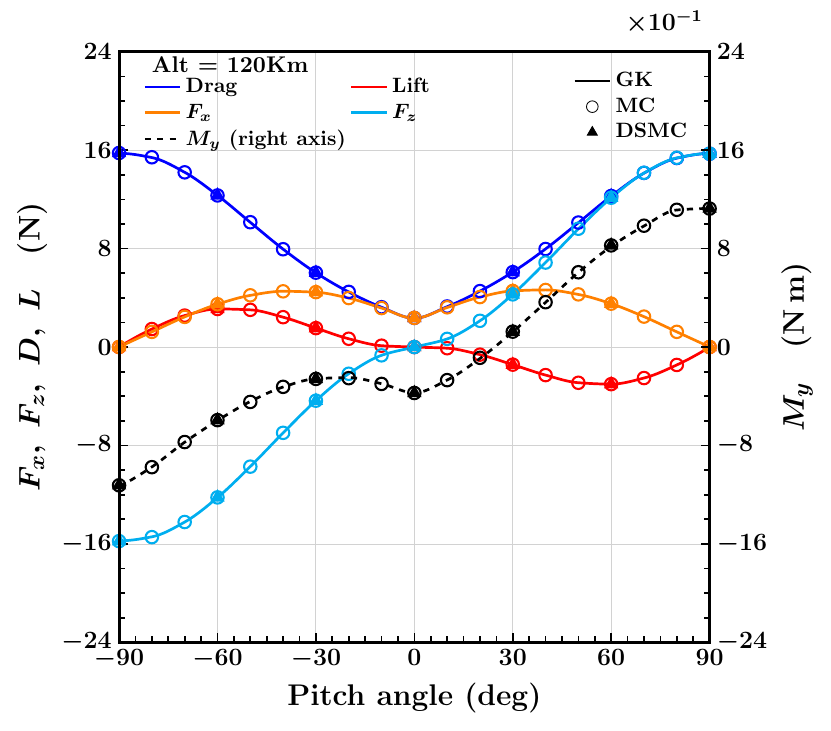}}
    \hfill
    \subfloat[150~km]{\includegraphics[width=0.48\textwidth]{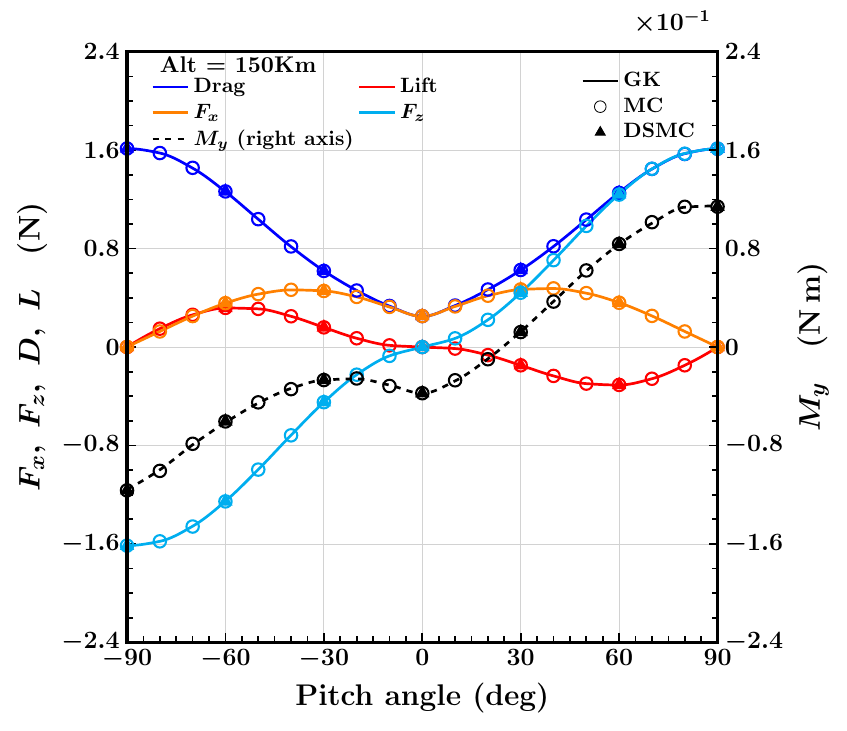}}
    \\[3pt]
    \subfloat[180~km]{\includegraphics[width=0.48\textwidth]{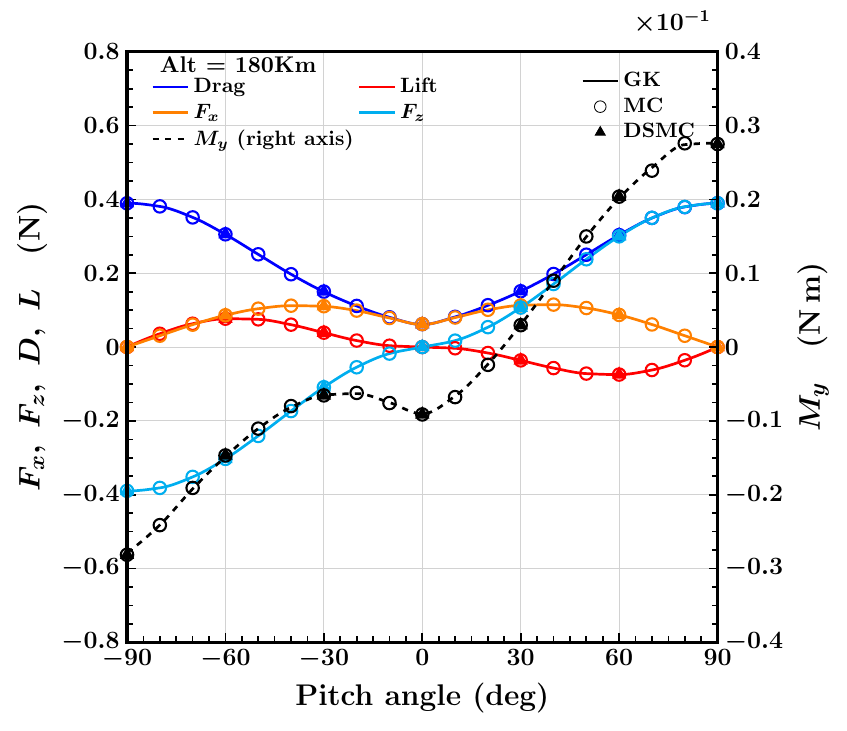}}
    \hfill
    \subfloat[240~km]{\includegraphics[width=0.48\textwidth]{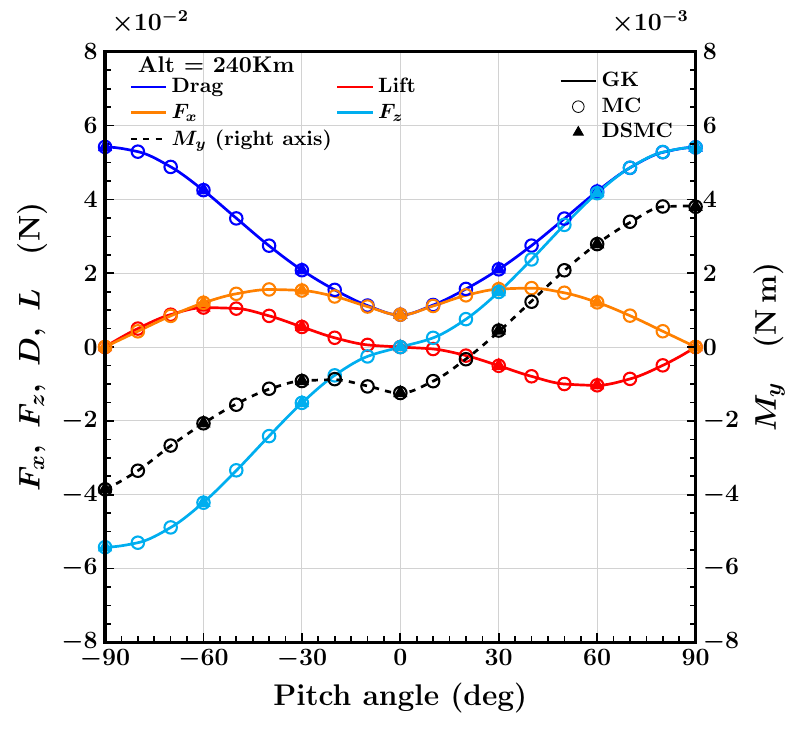}}
    \\[3pt]
    \subfloat[300~km]{\includegraphics[width=0.48\textwidth]{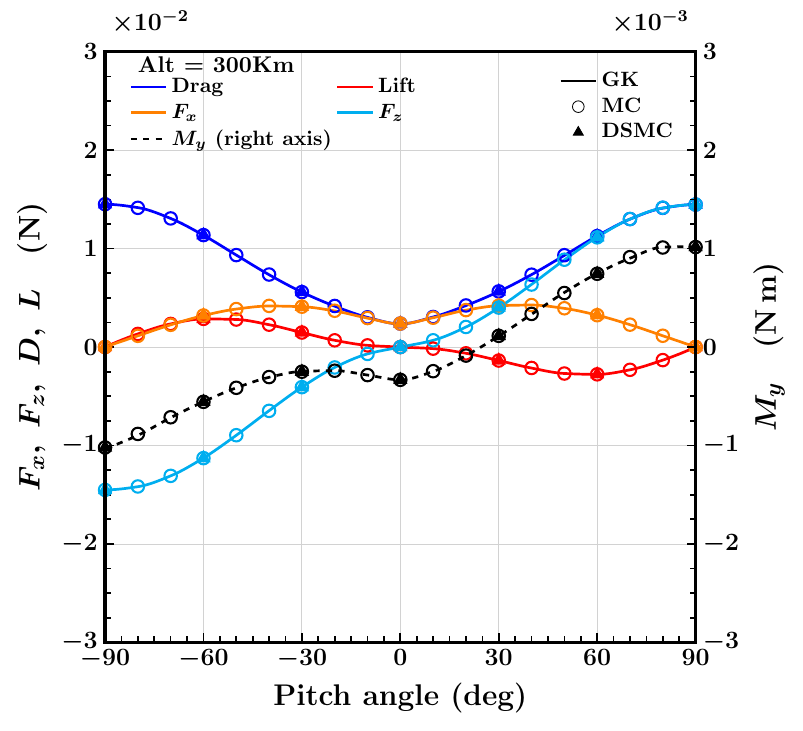}}
    \hfill
    \subfloat[Total heat flux]{\includegraphics[width=0.45\textwidth]{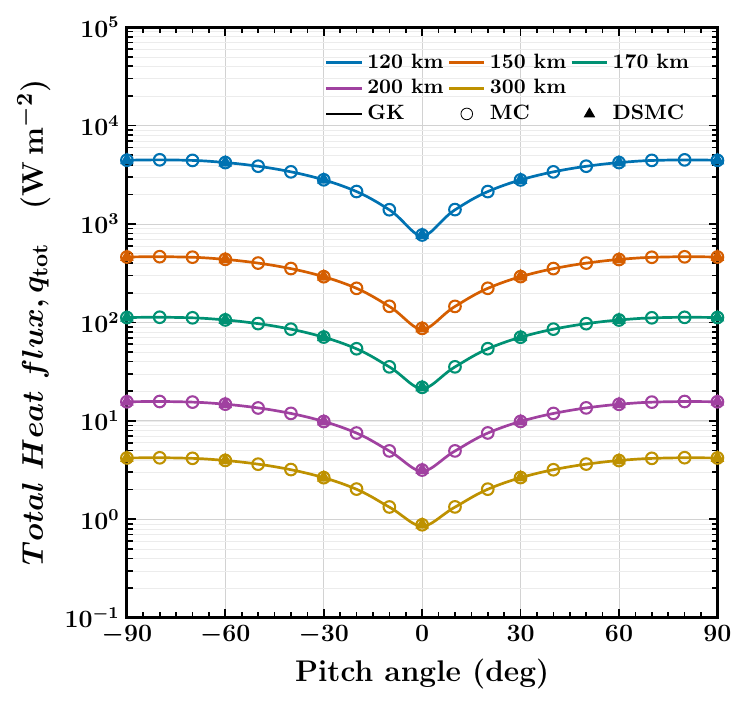}}
    \caption{Pitch-angle dependence of the aerodynamic loads and heating for the complex-satellite CS1 cases.}
    \label{fig:complex_pitch_forces}
\end{figure*}

The complex-satellite pitch sweeps from 120 to 300~km show a decrease in forces and moments with altitude (Fig.~\ref{fig:complex_pitch_forces}). At a given altitude, pitching changes the inward normal velocity on the array and appendages, as well as the fraction of each surface that molecules can reach. The drag is lowest near the nominal attitude and increases as more of the geometry presents a projected area to the incoming flow. The body-normal force $F_z$ and lift change sign as the direction of normal loading reverses. The axial force $F_x$ retains its sign in the plotted convention.

The pitching moment has a less regular angular dependence because the load can shift between surfaces with different lever arms as they become exposed or shielded. The heating response also depends on which surfaces receive the incident kinetic energy and how much energy leaves with the reflected molecules. GK--TAMV and MC recover similar integrated trends over the pitch range.

\subsubsection{Combined Altitude and Yaw-Angle Effects}
\label{sec:complex_yaw}

\begin{figure*}[htbp]
    \centering
    \subfloat[120~km]{\includegraphics[width=0.48\textwidth]{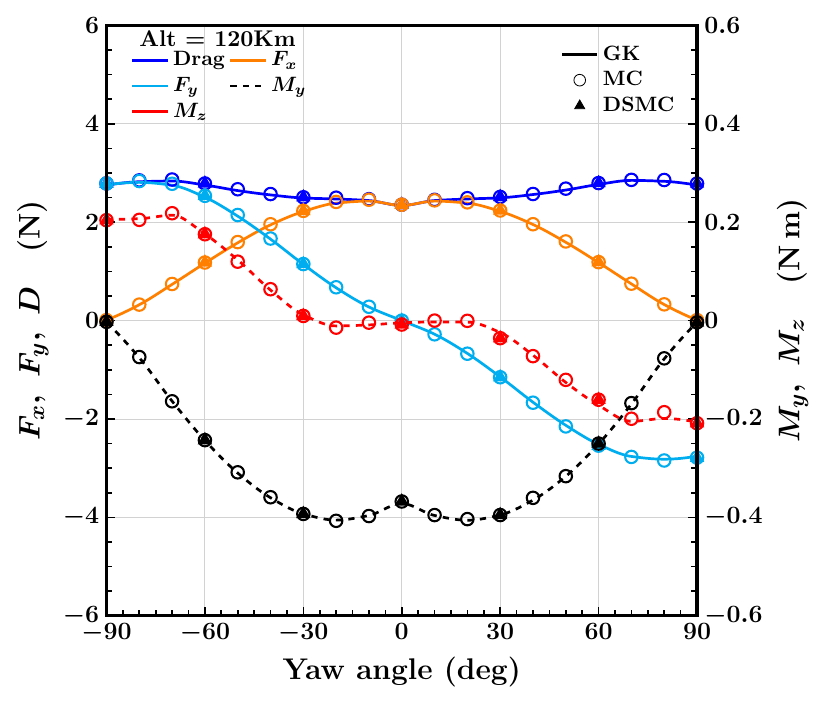}}
    \hfill
    \subfloat[180~km]{\includegraphics[width=0.48\textwidth]{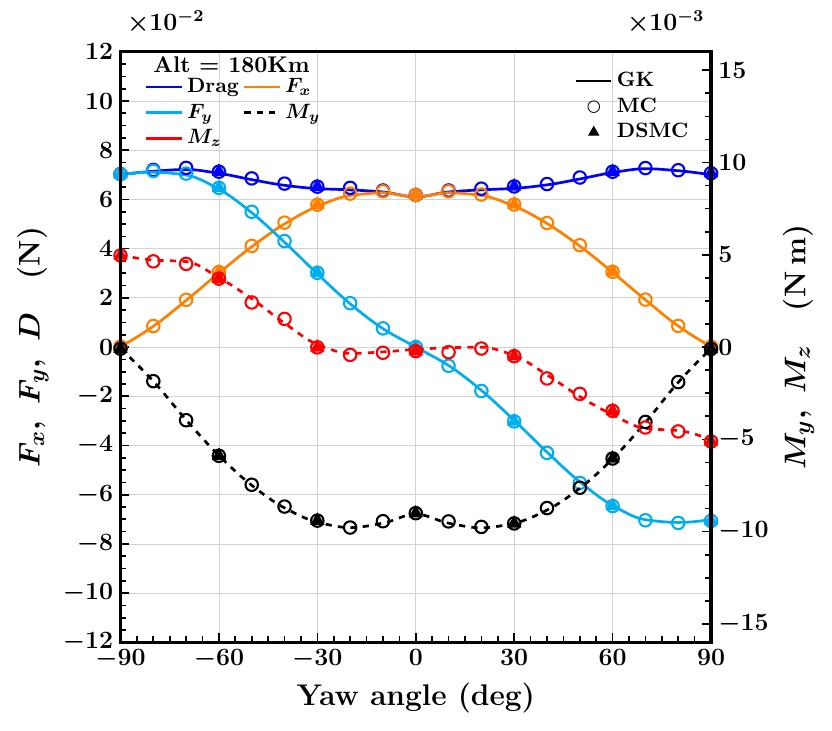}}
    \\[3pt]
    \subfloat[240~km]{\includegraphics[width=0.48\textwidth]{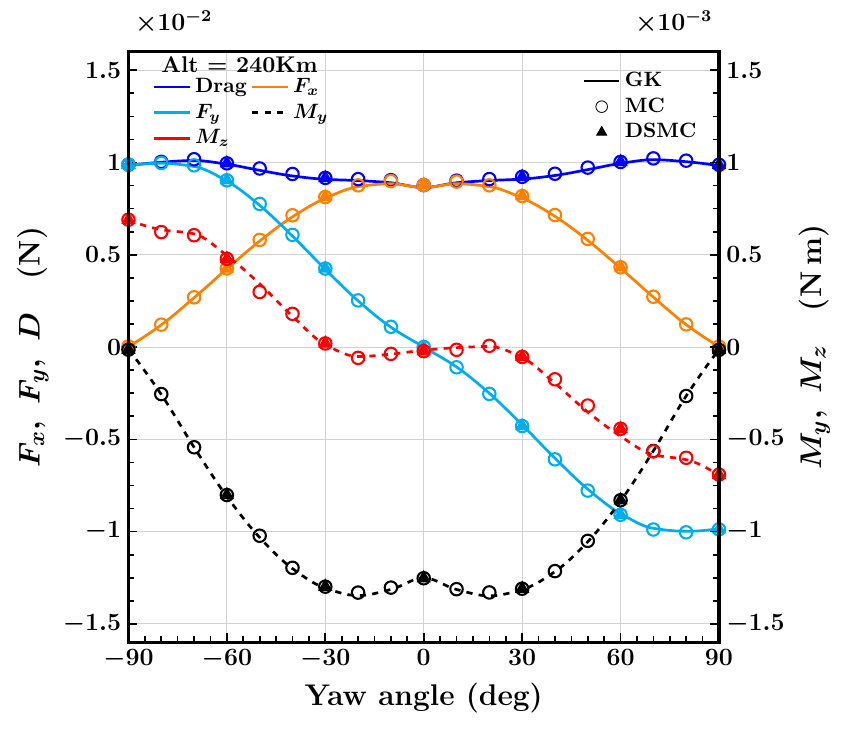}}
    \hfill
    \subfloat[Total heat flux]{\includegraphics[width=0.45\textwidth]{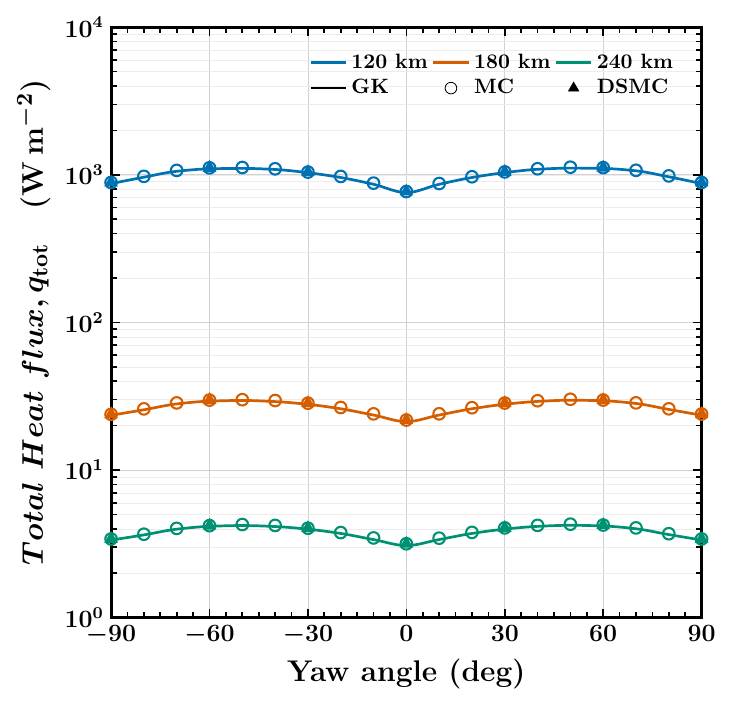}}
    \caption{Yaw-angle dependence of the aerodynamic loads and heating for the complex-satellite CS2 cases at 120, 180, and 240~km.}
    \label{fig:complex_yaw_forces}
\end{figure*}

The complex-satellite yaw sweeps at 120, 180, and 240~km show a smaller change in drag than the pitch sweeps (Fig.~\ref{fig:complex_yaw_forces}). Yaw primarily redistributes the load between lateral surfaces. Accordingly, $F_y$ changes sign as the flow moves from one side to the other, and $F_x$ is approximately symmetric about zero yaw. The yawing moment depends on the spanwise position of the axial loads and the axial position of the lateral loads. Components can contribute moments of opposite sign even when their forces contribute to the same total drag.

At large yaw angles, some appendages approach grazing incidence or lie behind another component along the mean flow. Under these conditions the incident momentum flux depends on which molecular trajectories can reach each surface. GK--TAMV and MC show similar integrated trends, and the local effect of direction-dependent exposure is examined in Section~\ref{sec:complex_surface}.

\subsubsection{Accommodation-Coefficient Effects at 150~km}
\label{sec:complex_accommodation}

\begin{figure*}[htbp]
\centering
\includegraphics[width=0.32\textwidth]{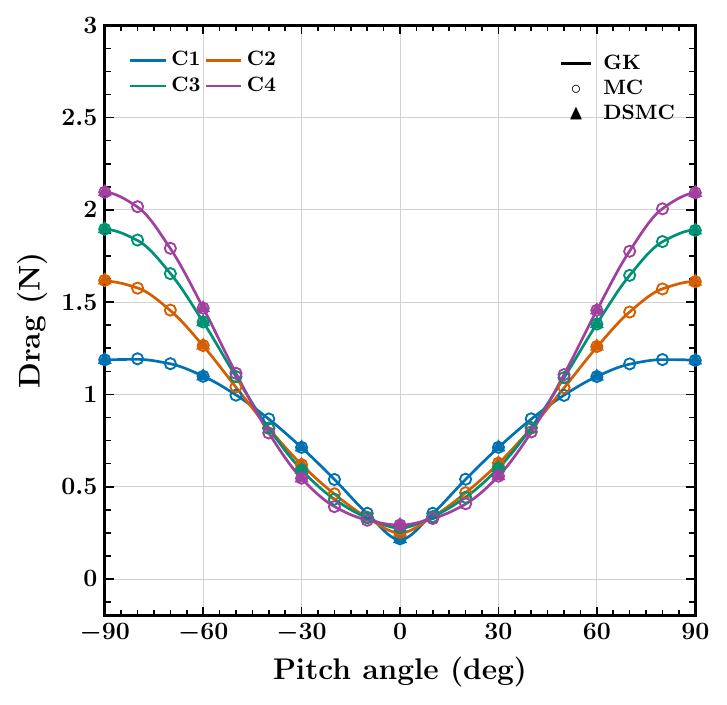}\hfill
\includegraphics[width=0.32\textwidth]{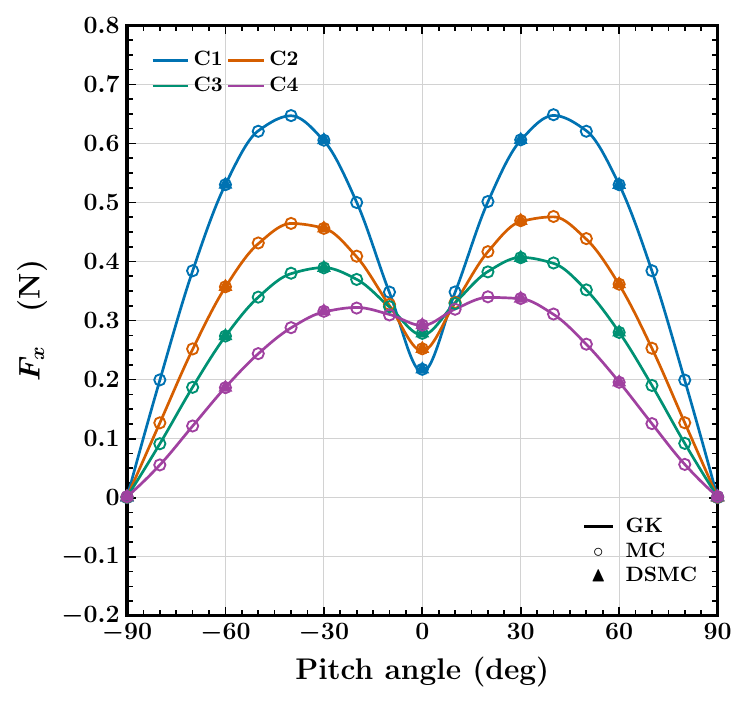}\hfill
\includegraphics[width=0.32\textwidth]{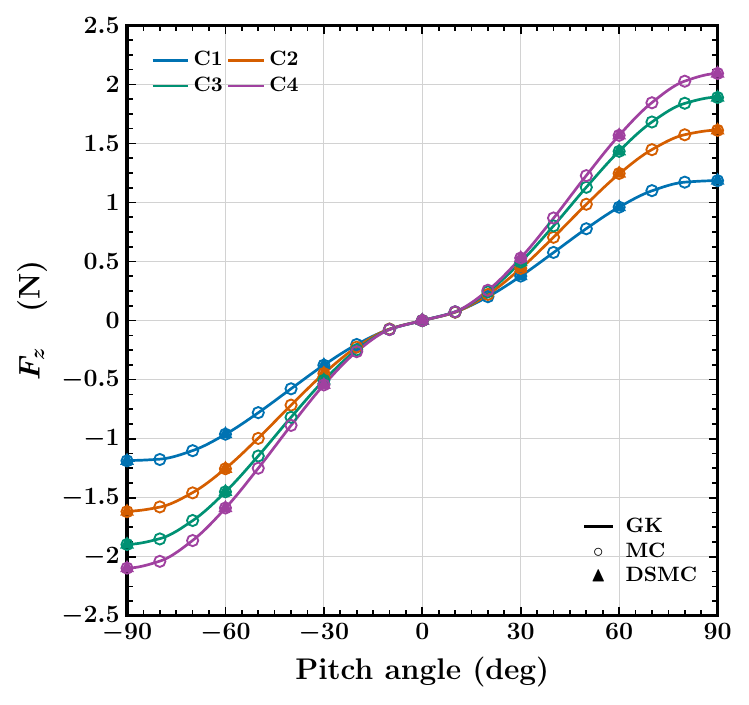}\\[3pt]
\includegraphics[width=0.32\textwidth]{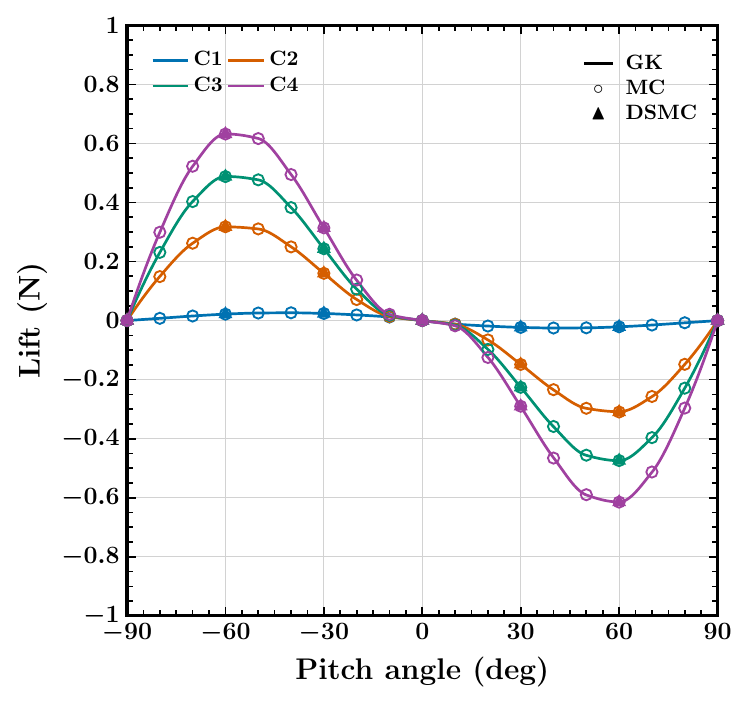}\hfill
\includegraphics[width=0.32\textwidth]{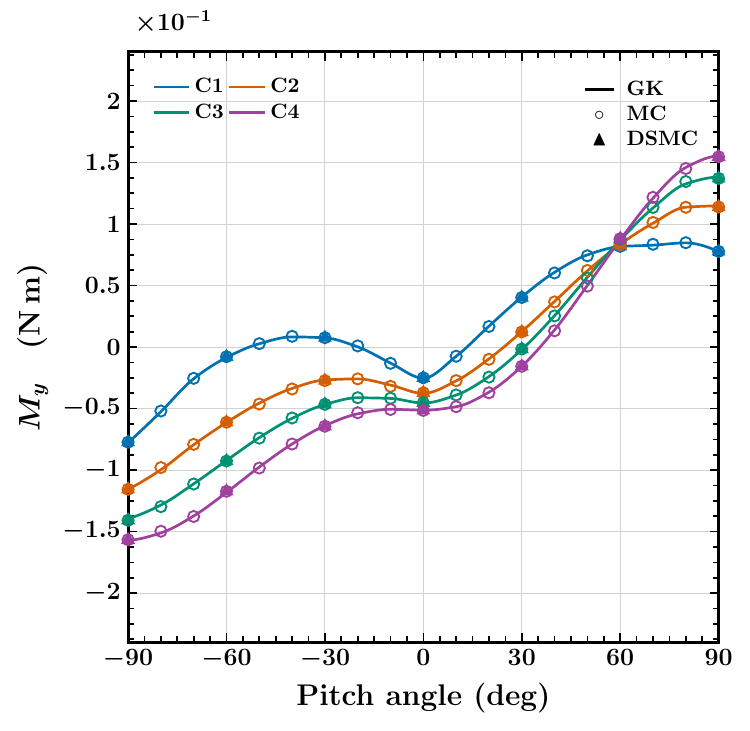}\hfill
\includegraphics[width=0.32\textwidth]{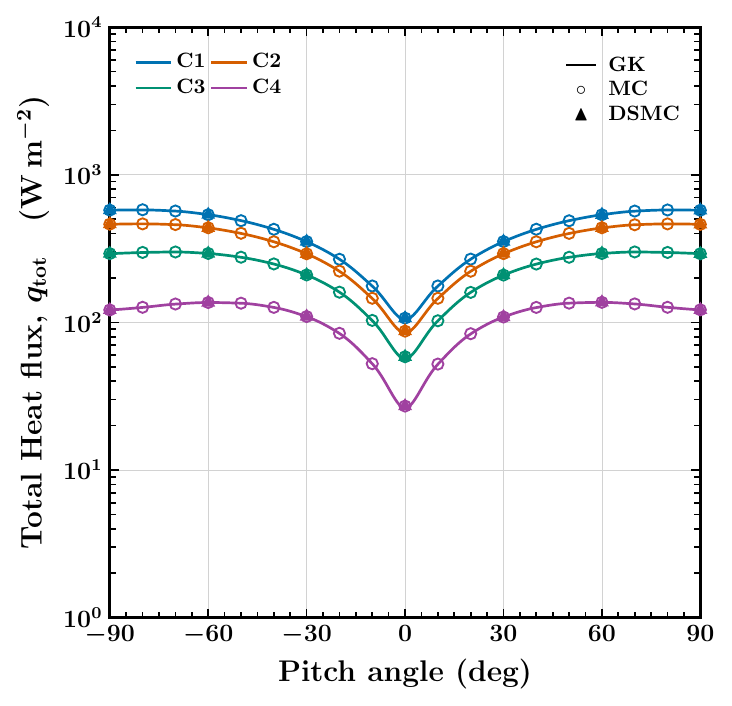}
\caption{CS3 accommodation sensitivity of the complex satellite at 150~km: (a) drag, (b) $F_x$, (c) $F_z$, (d) lift, (e) pitching moment $M_y$, and (f) total heat flux.}
\label{fig:complex_accommodation}
\end{figure*}

The CS3 calculations at 150~km use the same C1--C4 accommodation sets as the SLATS case (Section~\ref{sec:jaxa_geometry_cases}). Figure~\ref{fig:complex_accommodation} shows that the drag curves retain their broad pitch dependence and separate more at large pitch magnitudes. At a fixed attitude, the geometry and incidence are common to all four sets. The changed normal and tangential momentum transfer therefore alters the traction on the same exposed surfaces. The effects on lift and pitching moment depend on the orientation and lever arm of those surfaces.

The heating curves are also sensitive to accommodation through the reflected translational energy, with the internal-energy accommodation fixed at $\alpha_E=1$. GK--TAMV closely follows the MC trends.

\subsubsection{Surface-Pressure Distribution and Molecular Shadowing}
\label{sec:complex_surface}

\begin{figure*}[htbp]
\centering

\subfloat[GK--TAMV]{%
\includegraphics[width=0.48\textwidth]
{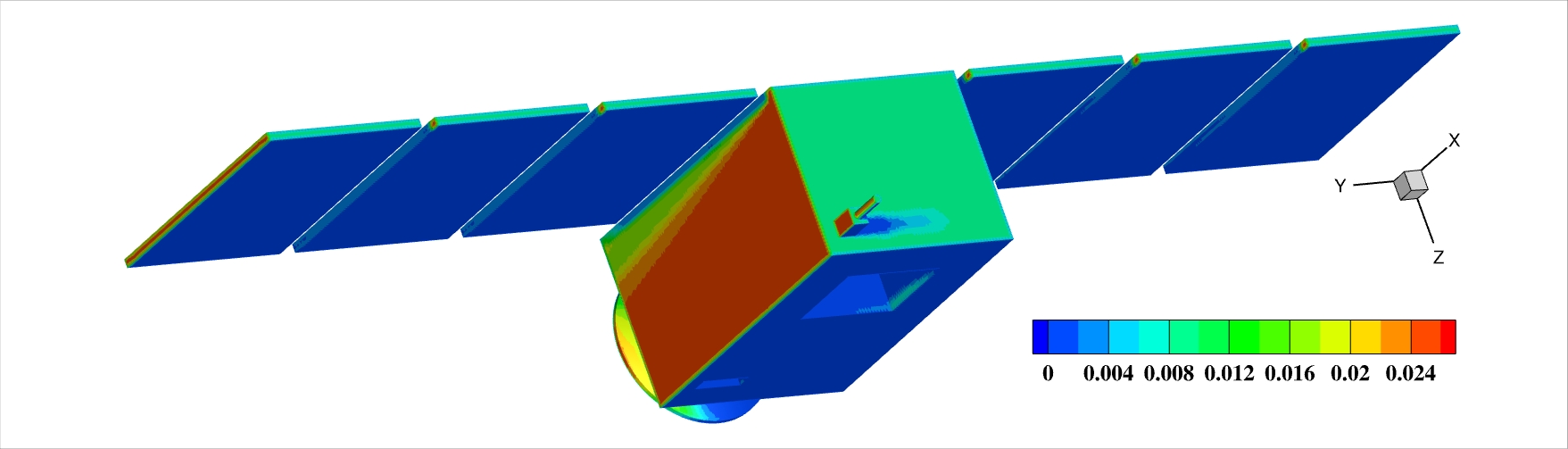}
}
\hfill
\subfloat[GK conventional]{%
\includegraphics[width=0.48\textwidth]
{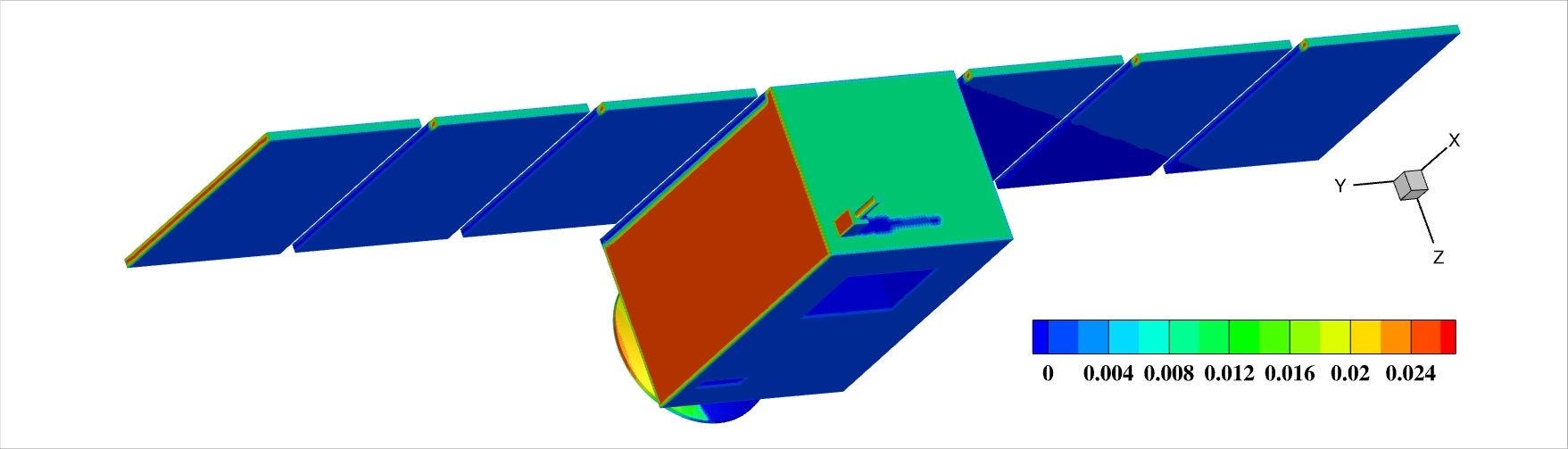}
}

\vspace{2mm}

\subfloat[Monte Carlo]{%
\includegraphics[width=0.48\textwidth]
{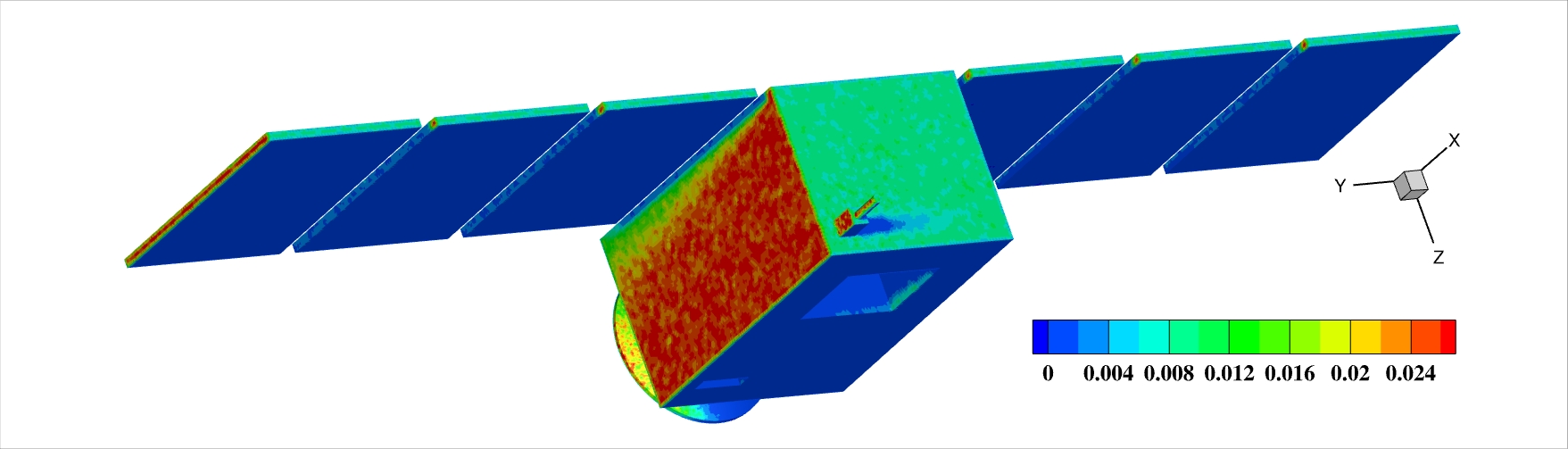}
}
\hfill
\subfloat[DSMC]{%
\includegraphics[width=0.48\textwidth]
{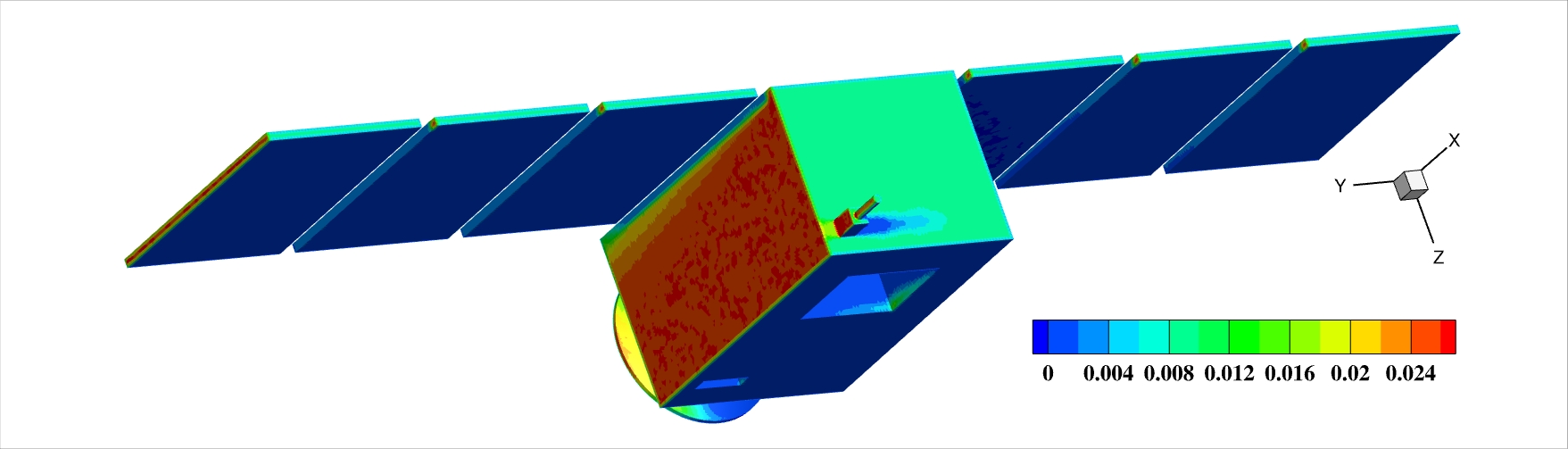}
}

\caption{Complex-satellite surface pressure at 180~km and $\psi=-60^\circ$ from (a) GK--TAMV, (b) conventional GK, (c) MC, and (d) DSMC. The main departures between the GK maps occur along component shadow boundaries and near grazing incidence.}
\label{fig:complex_pressure_comparison}
\end{figure*}

At 180~km and a yaw angle of $-60^\circ$, all four methods in Fig.~\ref{fig:complex_pressure_comparison} place the highest pressure on the exposed upstream face of the bus, where the incident molecules have a large inward normal velocity and direct access to the surface. The arrays and downstream surfaces are more lightly loaded because of weak normal incidence, geometric shielding, or both.

The two GK maps differ most along the upper and lateral bus faces and around the appendages. In these regions a small change in the incident velocity direction can move a trajectory across a shadow boundary. TAMV accounts for the fraction of the incident velocity distribution that reaches each panel, with higher inward normal speeds carrying more weight in the impact rate. The momentum transferred by these impacts yields finite pressure in some regions that conventional GK classifies as more strongly shielded. MC and DSMC show a similar redistribution of pressure, with visible fluctuations between panels on the loaded face.

\subsection{Limitations and Future Work}
\label{sec:limitations_future_work}

HACS computes aerodynamic forces, moments, and heating for the spacecraft cases considered here down to 120~km, the lowest altitude examined. Its applicability at a given altitude depends on the Knudsen number based on the local atmospheric state and a characteristic spacecraft length. HACS assumes free-molecular transport and accounts for the first surface impact. As intermolecular collisions become important, or as reflected molecules strike another part of the spacecraft, those processes can change the predicted loads. The SPARTA comparisons at selected conditions provide a check on the reported cases.

A next step is to use the gas-kinetic solver to generate training data rapidly along prescribed orbital trajectories. At each sampled time, the spacecraft location and the solar and geomagnetic inputs can be passed to NRLMSIS to obtain the local neutral-species number densities and temperature~\cite{Emmert2021NRLMSIS}. The trajectory, assumed atmospheric motion, and spacecraft attitude determine the relative-flow velocity in the body frame. Together with wall temperature and accommodation parameters, these quantities define each HACS case. Repeating the calculation along the orbit yields a database of forces, moments, and heating rates for that geometry.

A neural-network surrogate trained on these databases could then predict loads along new orbital and attitude histories within the range of conditions covered by the training data. Evaluation on complete orbital sequences held out from training would test how well the surrogate reproduces the variation in loads along a new trajectory. Selected SPARTA cases near the onset of transitional flow could help identify where the free-molecular training data cease to provide a suitable reference.

\section{Conclusion}\label{S:5}

HACS combines gas-kinetic flux integration and first-hit Monte Carlo sampling for free-molecular aerodynamics and heating. The two solvers use the same species-resolved freestream inputs and surface laws. Separating incident and reflected terms makes it possible to track how molecular access and surface scattering contribute to the net transfer.

The flat-plate comparisons showed close agreement between GK and MC for the incident coefficients and for Maxwell and CLL reflection. In the independent CLL comparison with SPARTA, the maximum relative drag differences were $0.863\%$ for CLL-exact and $0.876\%$ for MC. The corresponding lift differences were at most $0.990\%$ and $0.938\%$ when the symmetry point at normal incidence was excluded. In the conventional GK flat-plate test, evaluating the CLL normal moment at one representative incident speed gave its largest relative pressure difference of about $4.08\%$ near grazing incidence, where the absolute coefficient difference was $1.20\times10^{-4}$.

For the two spacecraft, increasing altitude reduced the load and heating scale, while attitude changed their distribution. Bulk-direction visibility gave nearly the same integrated drag as TAMV in the SLATS pitch sweep at 200~km. The yaw and local pressure comparisons showed the clearest benefit of TAMV near grazing surfaces and component shadow boundaries. Changing the CLL normal and tangential accommodation also changed net heating at fixed internal-energy accommodation, through the energy carried by the reflected translational motion.

These conclusions apply to first-hit, free-molecular transport. Intermolecular collisions and subsequent surface impacts can affect the loads when mean free paths are comparable to spacecraft dimensions or when a geometry traps reflected molecules. The SPARTA comparisons benchmark the conditions sampled here.


\section*{Author Declarations}

\subsection*{Conflict of Interest}
No conflict of interest

\subsection*{Author Contributions}

\textbf{Vignesh Ram Petha Sethuraman:}
Conceptualization, Formal analysis, Investigation, Methodology,
Writing -- original draft.

\textbf{Wu Kexin:}
Methodology, Validation, Writing -- review \& editing.

\textbf{S Ajith Kumar:}
Investigation, Supervision, Writing -- review \& editing.

\section*{Data Availability}
The data that support the findings of this study are available from the corresponding author upon reasonable request. Precompiled HACS executables for mesh preprocessing (Prepro) and gas--surface interaction (GSI) calculations, together with instructions for setting up and running a case, are available at \url{https://github.com/rampsv/HACS_GSI}.

\appendix

\section{Incident Free-Molecular Fluxes}
\label{app:incident_flux}

\subsection{Half-range velocity moments}

Consider an unobstructed panel and one species $s$. Use the outward normal $\mathbf n$, inward molecular speed $w>0$, and tangential velocity $\mathbf v_t=\mathbf v+w\mathbf n$ defined in Section~\ref{sec:incident_flux}. The Maxwellian in Eq.~\eqref{eq:drifting_maxwellian} factors in these coordinates as
\begin{equation}
 f_s(\mathbf v)
 = n_s\frac{\beta_s}{\sqrt{\pi}}e^{-(\beta_s w-s_n)^2}
       \frac{\beta_s^2}{\pi}e^{-\beta_s^2|\mathbf v_t-\mathbf V_T|^2}.
 \label{eq:app_maxwellian}
\end{equation}
The two tangential components have means given by $\mathbf V_T$ and variance $1/(2\beta_s^2)$ in each component. The tangential density integrates to unity. The flux through a unit area into the wall is weighted by $w$, as in classical free-molecular theory~\cite{sentman_1961,Bird1994}. For $x=\beta_s w$ define
\begin{equation}
 H_j=\int_0^\infty x^j e^{-(x-s_n)^2}\,dx,
 \qquad M_j=\frac{H_j}{\sqrt{\pi}\beta_s^j}.
 \label{eq:app_half_range}
\end{equation}
Writing $x=s_n+(x-s_n)$ and using $(x-s_n)e^{-(x-s_n)^2}=-\tfrac12\,d[e^{-(x-s_n)^2}]/dx$ gives $H_1=s_nH_0+E_n/2$. For $j\ge2$ the boundary term vanishes, giving
\begin{equation}
 H_j=s_n H_{j-1}+\frac{j-1}{2}H_{j-2}.
 \label{eq:app_recurrence}
\end{equation}
With $E_n$ and $R_n$ defined in Section~\ref{sec:maxwell_gsi}, the moments needed here are
\begin{align}
 H_0 &= \frac{\sqrt{\pi}}2 R_n,
 & H_1 &= \frac{E_n}{2}+s_nH_0=\frac{A_n}{2},\nonumber\\
 H_2 &= \frac{s_nE_n}{2}+\left(s_n^2+\frac12\right)H_0,
 & H_3 &= \frac{(s_n^2+1)E_n}{2}+\left(s_n^3+\frac32s_n\right)H_0.
 \label{eq:app_moments}
\end{align}
These expressions retain the sign of $s_n$. At $s_n=0$, for example, thermal motion still gives $\Gamma_s=n_s/(2\sqrt{\pi}\beta_s)$.

\subsection{Momentum and energy delivered to the panel}

Integrating the molecular rate and the rate-weighted momentum yields
\begin{align}
 \Gamma_s &= \int_{w>0} w f_s(\mathbf v)\,d^3v=n_sM_1,\\
 \mathbf f_{i,s}
 &=\int_{w>0} w f_s(\mathbf v)m_s\mathbf v\,d^3v
 =-\rho_s M_2\mathbf n+\rho_s M_1\mathbf V_T .
 \label{eq:app_incident_momentum}
\end{align}
The wall-normal traction is compressive. Its magnitude and the tangential traction give Eqs.~\eqref{eq:incident_pressure} and \eqref{eq:incident_shear}. The tangential mean squared speed is $|\mathbf V_T|^2+\beta_s^{-2}$. Integrating $e_i$ against the incident flux therefore gives
\begin{align}
 q_{i,s,\mathrm{tr}}
 &=\frac{\rho_s}{2}\bigl[M_3+M_1(|\mathbf V_T|^2+\beta_s^{-2})\bigr],\nonumber\\
 q_{i,s,\mathrm{int}} &= \Gamma_s\frac{\zeta_s}{2}k_BT_\infty .
 \label{eq:app_incident_energy}
\end{align}
Their sum is Eq.~\eqref{eq:incident_energy}. Using $n_s k_BT_\infty=\rho_s/(2\beta_s^2)$, the conventional GK normalizations give
\begin{align}
 C_{p,i}
 &=\frac{2H_2}{\sqrt{\pi}}
  =\frac{s_n E_n}{\sqrt{\pi}}+\left(s_n^2+\frac12\right)R_n,\\
 C_{t,i}
 &=\frac{2s_t H_1}{\sqrt{\pi}}=\frac{s_t A_n}{\sqrt{\pi}},\\
 C_{h,i,\mathrm{tr}}
 &=\frac{H_3+(s_t^2+1)H_1}{\sqrt{\pi}}\nonumber\\
 &=\frac{(S_s^2+2)E_n+\sqrt{\pi}s_nR_n(S_s^2+5/2)}{2\sqrt{\pi}},\\
 C_{h,i,\mathrm{int}}
 &=\frac{\zeta_sH_1}{2\sqrt{\pi}}=\frac{\zeta_sA_n}{4\sqrt{\pi}} .
\end{align}
All expressions in this section are per unit area of unobstructed surface. Conventional GK applies $\chi_a^{\mathrm{conv}}$ as in Eqs.~\eqref{eq:element_force} and \eqref{eq:species_sum}.

\section{Reflected Momentum and Energy for Maxwell and CLL Scattering}
\label{app:cll_derivation}

For a prescribed incident velocity $\mathbf v_i=-w\mathbf n+\mathbf v_{i,t}$, write the reflected velocity as $\mathbf v_r=v_{r,n}\mathbf n+\mathbf v_{r,t}$, with $v_{r,n}\ge0$. An overbar denotes a conditional mean over the reflected state at fixed $\mathbf v_i$. The one outgoing molecule per incident molecule assumption gives the reflected wall traction and outgoing energy flux
\begin{equation}
 \mathbf f_{r,s}=-m_s\int_{w>0} w f_s(\mathbf v_i)\overline{\mathbf v}_r\,d^3v_i,
 \qquad
 q_{r,s}=\int_{w>0} w f_s(\mathbf v_i)\overline e_r\,d^3v_i .
 \label{eq:app_ref_integrals}
\end{equation}
Thus $p_{r,s}=-\mathbf f_{r,s}\cdot\mathbf n$ is positive in compression and the net energy flux into the wall is $q_{i,s}-q_{r,s}$.

\subsection{Maxwell conditional moments}

Specular reflection gives $v_{r,n}=w$, $\mathbf v_{r,t}=\mathbf v_{i,t}$, and $e_r=e_i$. For diffuse emission, the normalized outgoing normal density is $2\beta_w^2 u\exp(-\beta_w^2u^2)$ for $u=v_{r,n}>0$~\cite{Maxwell1879,Bird1994}. It has mean $\sqrt{\pi}/(2\beta_w)$ and second moment $\beta_w^{-2}$. The two tangential components have zero mean and together contribute $\beta_w^{-2}$ to the mean squared speed. The resulting diffuse mean energy is $e_w=(\zeta_s+4)k_BT_w/2$. With the GK accommodation rules of Section~\ref{sec:maxwell_gsi},
\begin{align}
 \overline{\mathbf v}_r^{M}
 &= (1-\sigma_T)\mathbf v_{i,t}
 +\left[(1-\sigma_N)w+\sigma_N\frac{\sqrt{\pi}}{2\beta_w}\right]\mathbf n,\\
 \overline e_r^{M} &= (1-\sigma_E)e_i+\sigma_E e_w .
 \label{eq:app_maxwell_means}
\end{align}
Substitution in Eq.~\eqref{eq:app_ref_integrals} gives
\begin{align}
 p_{r,s}^M&=(1-\sigma_N)p_{i,s}+\sigma_N m_s\Gamma_s\frac{\sqrt\pi}{2\beta_w},\\
 \boldsymbol\tau_{r,s}^M&=-(1-\sigma_T)\boldsymbol\tau_{i,s},\\
 q_{r,s}^M&=(1-\sigma_E)q_{i,s}+\sigma_E\Gamma_s e_w.
\end{align}
Using $H_1=A_n/2$ and $\beta_s/\beta_w=\sqrt{t_r}$ recovers Eqs.~\eqref{eq:maxwell_pressure_ref}--\eqref{eq:maxwell_heat_ref}. A single Maxwell specular--diffuse draw reproduces these conditional moments when the three accommodation parameters are equal.

The diffuse sampling also gives the signed projection used in Eq.~\eqref{eq:diffuse_signed_angle_pdf}. After integrating over the out-of-plane component, let $u=\beta_w v_{r,n}$ and $y=\beta_w\mathbf v_r\cdot\mathbf t$. Their joint density is $2u e^{-(u^2+y^2)}/\sqrt\pi$ for $u>0$. With $u=R\cos\theta_r^s$ and $y=R\sin\theta_r^s$, the Jacobian is $R$. Integration over $R>0$ gives $P(\theta_r^s)=\tfrac12\cos\theta_r^s$ on $(-\pi/2,\pi/2)$.

\subsection{CLL kernel and conditional moments}

The shifted Gaussian representation in Eq.~\eqref{eq:cll_normal_sampling} has mean $(\mu,0)$ in two dimensions and component variance $\alpha_N/2$, where $\mu=\sqrt{1-\alpha_N}\,c_i$. Transforming to polar coordinates and integrating the angle using the integral representation of $I_0$ gives Eq.~\eqref{eq:cll_kernel}~\cite{CercignaniLampis1971,Lord1991,DLMFBessel}. To evaluate its first moment, use
\begin{equation}
 \mathcal J(a,b)=\int_0^\infty e^{-a c^2}I_0(bc)\,dc
 =\frac{\sqrt\pi}{2\sqrt a}e^{b^2/(8a)}I_0\!\left(\frac{b^2}{8a}\right),
 \qquad a>0 .
\end{equation}
The identity follows by inserting $I_0(bc)=\pi^{-1}\int_0^\pi e^{bc\cos\phi}\,d\phi$. Pairing angles $\phi$ and $\pi-\phi$ converts the half-line Gaussian integral to half its full-line value. Completing the square gives the remaining angular average of $\exp[b^2\cos^2\phi/(4a)]$, which reduces to the displayed Bessel function by $\cos^2\phi=(1+\cos2\phi)/2$.

The integral for $\overline c_r$ contains $c^2 e^{-ac^2}I_0(bc)$. At fixed $b$, differentiating $\mathcal J$ with respect to $a$ and using $I_0'=I_1$~\cite{DLMFBessel} gives
\begin{equation}
 -\frac{\partial\mathcal J}{\partial a}
 =\frac{\sqrt\pi}{4a^{3/2}}e^z
 \bigl[(1+2z)I_0(z)+2zI_1(z)\bigr],
 \qquad z=\frac{b^2}{8a}.
\end{equation}
Substitute $a=1/\alpha_N$ and $b=2\mu/\alpha_N$ after differentiating. Multiplying by the kernel prefactor $2e^{-\mu^2/\alpha_N}/\alpha_N$ and dividing by $\beta_w$ gives
\begin{equation}
 \overline v_{r,n}=\frac{\sqrt{\pi\alpha_N}}{2\beta_w}\mathcal B(z),
 \qquad z=\frac{(1-\alpha_N)(\beta_w w)^2}{2\alpha_N}.
 \label{eq:app_cll_normal_mean}
\end{equation}
At $\alpha_N=0$ the normal speed is $w$. At $\alpha_N=1$ the normal distribution is diffuse and $\mathcal B(0)=1$.

The expectation of the squared Gaussian norm is $\mu^2+\alpha_N$. Applying the same mean and variance rule to the two tangential components in Eq.~\eqref{eq:cll_tangential_sampling} yields
\begin{align}
 \overline{\mathbf v}_{r,t}&=(1-\alpha_T)\mathbf v_{i,t},\\
 \overline{v_{r,n}^{,2}}&=(1-\alpha_N)w^2+\alpha_N\beta_w^{-2},\\
 \overline{|\mathbf v_{r,t}|^2}
 &=(1-\alpha_T)^2|\mathbf v_{i,t}|^2+a_T\beta_w^{-2}.
 \label{eq:app_cll_tangential_energy}
\end{align}
The conditional reflected velocity vector is $\overline v_{r,n}\mathbf n+\overline{\mathbf v}_{r,t}$. The outgoing mean total energy is
\begin{align}
 \overline e_r^{\mathrm{CLL}}
 &=\frac{m_s}{2}\left(\overline{v_{r,n}^{,2}}+\overline{|\mathbf v_{r,t}|^2}\right)
 +\overline e_{r,\mathrm{int}},\nonumber\\
 \overline e_{r,\mathrm{int}}
 &=\frac{\zeta_s k_B}{2}\left[(1-\alpha_E)T_\infty+\alpha_E T_w\right].
 \label{eq:app_cll_total_energy}
\end{align}

\subsection{Integration for conventional GK}

After tangential integration, the pressure integral in Eq.~\eqref{eq:app_ref_integrals} reduces to
\begin{equation}
 p_{r,s}^{\mathrm{CLL}}
 =\frac{m_s n_s}{\sqrt{\pi}\beta_s}
 \int_0^\infty x e^{-(x-s_n)^2}\overline v_{r,n}(x/\beta_s)\,dx .
\end{equation}
Substituting Eq.~\eqref{eq:app_cll_normal_mean} and dividing by $n_s k_BT_\infty$ gives Eq.~\eqref{eq:cll_exact_pressure}. The mean incident speed among wall impacts is $\int w^2 f_s\,d^3v/\int w f_s\,d^3v=M_2/M_1$, with both integrals over $w>0$. Evaluating the conditional mean once at that speed gives $p_{r,s}^{\mathrm{fast}}=m_s\Gamma_s\overline v_{r,n}(M_2/M_1)$, which reduces to Eq.~\eqref{eq:cll_fast_pressure}. At full normal accommodation the conditional mean is independent of $w$, so both expressions reduce to the diffuse pressure. For zero normal accommodation both return the incident pressure in the specular limit.

Integrating the tangential first moment gives $\boldsymbol\tau_{r,s}^{\mathrm{CLL}}=-(1-\alpha_T)\boldsymbol\tau_{i,s}$. The second moments give
\begin{align}
 q_{r,s,\mathrm{tr}}^{\mathrm{CLL}}
 &=\frac{\rho_s}{2}\Bigl[(1-\alpha_N)M_3
 +(1-\alpha_T)^2 M_1(|\mathbf V_T|^2+\beta_s^{-2})\nonumber\\
 &\hspace{8em} +(\alpha_N+a_T)M_1\beta_w^{-2}\Bigr],\\
 q_{r,s,\mathrm{int}}^{\mathrm{CLL}}
 &=\Gamma_s\overline e_{r,\mathrm{int}}.
\end{align}
Dividing by $n_s k_BT_\infty/\beta_s$ gives Eqs.~\eqref{eq:cll_heat_trans} and \eqref{eq:cll_heat_internal}. The energy calculation uses these second moments in both conventional pressure branches.

\section{Velocity-Dependent Visibility and Monte Carlo Estimators}
\label{app:mc_estimators}

\subsection{Gauss--Hermite integration of panel impact rates}
\label{app:tamv_derivation}

For GK--TAMV, consider an incoming molecular property $g_s(\mathbf v)$ on panel $a$. It may be scalar or vector valued. A reflected property enters through its conditional mean from Appendix~\ref{app:cll_derivation}. The panel integral is
\begin{equation}
 \mathcal R_{a,s}[g_s]=A_a\int f_s(\mathbf v)
 [-\mathbf v\cdot\mathbf n_a]_+\chi_{a,s}(\mathbf v)g_s(\mathbf v)\,d^3v ,
 \label{eq:app_panel_rate}
\end{equation}
where $[y]_+=\max(y,0)$ and $\chi_{a,s}(\mathbf v)$ is the exposed fraction for that direction, based on the panel's BVH visibility rule. Changing variables to $\mathbf X=\beta_s(\mathbf v-\mathbf U_\infty)$ gives $f_s\,d^3v=n_s\pi^{-3/2}e^{-|\mathbf X|^2}\,d^3X$. The tensor-product physicists' Gauss--Hermite rule for this weight~\cite{DLMFQuadrature} gives
\begin{equation}
 \mathcal R_{a,s}[g_s]\simeq
   \sum_{q:\,v_{n,a,s,q}>0}\dot N_{a,s,q}\,g_s(\mathbf v_{s,q}),
\end{equation}
with $\dot N_{a,s,q}$ from Eq.~\eqref{eq:tamv_rate}. Taking $g_s$ as $m_s\mathbf v$, $-m_s\overline{\mathbf v}_r$, $e_i$, and $\overline e_r$ yields Eq.~\eqref{eq:tamv_exchange} after summation over species. Taking $g_s=1$ with the actual exposed fractions and again with all fractions set to one gives the numerator and denominator of Eq.~\eqref{eq:tamv_visibility} after cancellation of $A_a$. This ratio measures number flux. Pressure and heating use their own velocity and energy weights inside the quadrature.

\subsection{Launch weighting and first-hit sampling}

For species $s$, let $\mathbf x$ be the launch location, $A_{\mathrm{launch}}$ the launch area, and $\mathbf n_{\mathrm{in}}(\mathbf x)$ the inward launch normal. The physical rate of molecules crossing $dA$ in velocity volume $d^3v$ is
\begin{equation}
 f_s(\mathbf v)[\mathbf v\cdot\mathbf n_{\mathrm{in}}(\mathbf x)]_+\,dA\,d^3v .
\end{equation}
The joint sampling density in Section~\ref{sec:mc_solver} is $f_s(\mathbf v)/(n_sA_{\mathrm{launch}})$. Dividing the physical rate density by this proposal density gives the rate factor $n_sA_{\mathrm{launch}}[\mathbf v\cdot\mathbf n_{\mathrm{in}}]_+$. The sample mean over $N_{\mathrm{MC}}$ attempts gives Eq.~\eqref{eq:mc_launch_rate}. Outward launches, misses, and rejected intersections contribute zero while remaining in the sample count.

Let $\mathbb I_{a,p}$ equal one when attempt $p$ yields an accepted first hit on panel $a$, and zero otherwise. Using the signed transfers in Eqs.~\eqref{eq:mc_momentum} and \eqref{eq:mc_energy}, the estimators for $j\in\{i,r\}$ are
\begin{align}
 \widehat{\mathbf F}_{j,a}&=\sum_s\sum_{p=1}^{N_{\mathrm{MC}}}\mathbb I_{a,p}W_p\Delta\mathbf p_{j,p},\\
 \widehat{\mathbf M}_{j,a}&=\sum_s\sum_{p=1}^{N_{\mathrm{MC}}}\mathbb I_{a,p}W_p
 (\mathbf x_{\mathrm{hit},p}-\mathbf x_0)\times\Delta\mathbf p_{j,p},\\
 \widehat{\dot Q}_{j,a}&=\sum_s\sum_{p=1}^{N_{\mathrm{MC}}}\mathbb I_{a,p}W_p\Delta E_{j,p}.
 \label{eq:app_mc_estimates}
\end{align}
Each term is defined as zero for rejected attempts. $W_p$ has units of molecules per second, so the three sums give force, moment, and power. The signed reflected terms add to the incident terms to yield net transfer. Division of the force and power by $A_a$ gives traction and heat flux. Summation over panels yields the spacecraft totals.

For independent draws on a launch surface that covers the contributing incoming trajectories, the expectation of each estimator equals the corresponding first-hit launch integral. Conditioning on an accepted incident state replaces the sampled outgoing momentum and energy by the conditional means of the MC scattering law. This connects the MC expectation to the GK integrands when the reflection moments agree. Finite GH order, panel-level visibility sampling, and the GK centroid approximation to the moment arm can still produce differences between GK--TAMV and MC.

\subsection{Sampling the number-flux visibility}

The separate MC visibility diagnostic uses velocity draws $\mathbf v_{s,p}\sim f_s/n_s$ and panel-based directional exposure. Define $w_{a,s,p}=[-\mathbf v_{s,p}\cdot\mathbf n_a]_+$. With $N_{v,s}$ diagnostic velocity draws for species $s$, the estimate is
\begin{equation}
 \widehat\chi_a^*=
 \frac{\sum_s(n_s/N_{v,s})\sum_p w_{a,s,p}\chi_{a,s}(\mathbf v_{s,p})}
      {\sum_s(n_s/N_{v,s})\sum_p w_{a,s,p}} .
\end{equation}
As in GK--TAMV, the limit of this ratio is the accessible fraction of incident number flux for the adopted panel-visibility rule. The first-hit load estimators in Eq.~\eqref{eq:app_mc_estimates} already account for geometric accessibility through $\mathbb I_{a,p}$.

\bibliography{HACS_references}
\end{document}